\documentclass[lettersize,journal]{IEEEtran}
\usepackage{amsmath,amsfonts}
\usepackage{algorithmic}
\usepackage{algorithm}
\usepackage{array}
\usepackage[caption=false,font=normalsize,labelfont=sf,textfont=sf]{subfig}
\usepackage{textcomp}
\usepackage{stfloats}
\usepackage{url}
\usepackage{verbatim}
\usepackage{graphicx}
\usepackage{cite}

\usepackage{hyperref}
\usepackage{harmony}
\usepackage{musicography}
\usepackage{cleveref}
\usepackage[table]{xcolor}
\usepackage[defaultlines=3,all]{nowidow}
\usepackage{makecell}

\newcommand{\chg}[2]{{\color{orange}$^{\textbf{old:}}$ #1} {\color{teal}$^{\textbf{new}:}$ #2}}
\renewcommand{\chg}[2]{{#2}}

\newcommand{\chgmin}[2]{{\color{orange}$^{\textbf{old:}}$ #1} {\color{teal}$^{\textbf{new}:}$ #2}}
\renewcommand{\chgmin}[2]{{#2}}

\crefname{section}{sec.}{sec.}
\Crefname{section}{Section}{Sections}
\crefname{subsection}{sec.}{sec.}
\Crefname{subsection}{Section}{Sections}
\crefname{paragraph}{sec.}{sec.}
\Crefname{paragraph}{Section}{Sections}
\crefname{figure}{fig.}{fig.}
\Crefname{figure}{Figure}{Figures}
\crefname{table}{tab.}{tab.}
\Crefname{table}{Table}{Tables}

\begin{document}

\title{Make the Unhearable Visible: Exploring Visualization for Musical Instrument Practice}

\author{Frank Heyen, Michael Gleicher~\IEEEmembership{Senior Member,~IEEE}, and Michael Sedlmair~\IEEEmembership{Member,~IEEE}
\thanks{This paper was produced by the IEEE Publication Technology Group. They are in Piscataway, NJ.}
\thanks{Manuscript received April 19, 2021; revised August 16, 2021.}%
\thanks{Frank Heyen and Michael Sedlmair are with Visualization Research Center (VISUS), University of Stuttgart. E-mail: frank.heyen@visus.uni-stuttgart.de, michael.sedlmair@visus.uni-stuttgart.de}%
\thanks{Michael Gleicher is with the Department of Computer Sciences, University of Wisconsin, Madison. E-mail: gleicher@cs.wisc.edu}%
}

\markboth{IEEE TRANSACTIONS ON VISUALIZATION AND COMPUTER GRAPHICS, VOL.~XX, NO.~X, XXXX~XXXX}%
{Shell \MakeLowercase{\textit{et al.}}: A Sample Article Using IEEEtran.cls for IEEE Journals}

\IEEEpubid{0000--0000/00\$00.00~\copyright~2021 IEEE}

\maketitle

\begin{abstract}
We explore the potential of visualization to support musicians in instrument practice through real-time feedback and reflection on their playing.
Musicians often struggle to observe \chg{the}{} patterns in their playing and interpret them with respect to their goals. 
Our premise is that these patterns can be made visible with interactive visualization: we can make the unhearable visible. 
However, understanding the design of such visualizations is challenging: the diversity of needs, including different instruments, skills, musical attributes, and genres, means that any single use case is unlikely to illustrate the broad potential and opportunities. 
To address this challenge, we conducted a design exploration \chg{study}{} where we created and iterated on 33 designs, each focusing on a subset of needs, for example, only one musical skill.
Our designs are grounded in our own experience as musicians and the ideas and feedback of 18 musicians with various musical backgrounds \chg{}{and we evaluated them with 13 music learners and teachers}.
This paper presents the results of our exploration, focusing on a few example designs as instances of possible instrument practice visualizations. 
From our work, we draw design considerations that contribute to future research and products for visual \chg{musical}{} instrument education.
Supplemental materials are available at \href{https://github.com/visvar/mila}{github.com/visvar/mila}.
\end{abstract}


\section{Introduction}

\begin{figure*}
  \centering
  \includegraphics[width=\linewidth, alt={Thumbnails conveying the diversity of our designs, grouped by category.}]{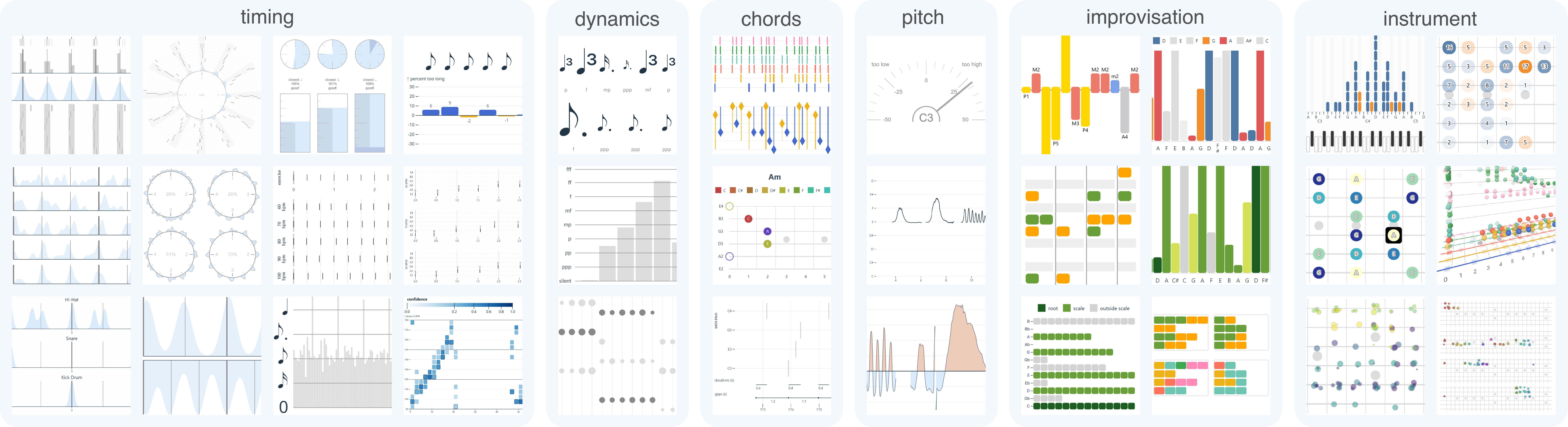}
  \caption{%
    \chg{%
  	We explored how visualization can support a broad range of musical instrument practice by making unhearable patterns visible.
    To this end, we created a collection of designs for different data attributes, musical instrument skills, and visual encodings.\\
    }{%
    An overview of the 33 designs we created to consider different data attributes, musical instrument skills, and visual encodings.
    They are categorized by the primary skill they support \chgmin{}{(see \Cref{sec:isolationofskills})}.
    Some of these designs are described in \chgmin{the paper}{\Cref{sec:design-exploration}}, all are provided in the supplement.%
    }
  }
  \label{fig:teaser}
\end{figure*}

\IEEEPARstart{D}{uring}
their practice, musicians acquire and improve a range of skills to play the instrument, including rhythmic timing, controlling the notes' pitch, and improvising.
Music education has traditionally included lessons and exercises.
Feedback is usually limited to listening, either the students listening to recordings 
\chg{}{of their practice}
or a teacher listening for them. 
However, listening is limited because some aspects are difficult to hear, especially for untrained musicians. 

The limitations of listening fall into two key categories. 
First, many important details are subtle and difficult for an untrained ear to identify but contribute to the overall experience. 
For example, an untrained ear may appreciate the ``groove'' of a song, but not be able to identify the subtle timing details that cause it. 
Second, many aspects of music relate to patterns over time, while hearing is instantaneous. 
While listening, the musicians must use their memory to build up distributions of instantaneous musical features over time.
For example, to get an overview of consistency in playing over multiple repetitions of an exercise, they would have to integrate a lot of information while playing.

These challenges with listening -- the need to identify details and summarize patterns and distributions -- are common strengths of visualization. Visualization can potentially make these difficult-to-hear things easy to see. 
Therefore, we propose to automatically create visualizations of musical practice data to support musicians in their practice. 
Many musicians already record data from their instruments. 
We propose tools to visualize this data. 
Our premise is that such visualizations, if designed properly, can support musicians in understanding their practice by making things difficult to hear easily visible.

Currently, automatic visual feedback is rarely used in music education, although we have anecdotal evidence such as using tuning devices to check intonation and looking at recorded audio waveforms or MIDI piano rolls to confirm playing on time.
The field of music visualization provides little consideration of 
practice support~\cite{wang2021soloist, yamabe2011feedback, heyen2022cellovis}, usually with a focus on narrow use cases.
Games~\cite{omeara2016rocksmith} and learning apps~\cite{ng2015easytolearnpiano} use visuals for instruction~\cite{skreinig2023guitarhero} and simple feedback~\cite{smith2008interactive, asahi2018toward:piano:support}, but lack detail and context that could support self-assessment.

\IEEEpubidadjcol

Unfortunately, music practice visualization is a challenging problem for which we have little guidance to create designs.
Musical practice involves diverse skills, instruments, and musicians. 
It further involves complex connections between measurable details, such as pitch and timing, and the perceived patterns that make the listening experience. 
Designs must bridge \chgmin{}{the gap} between the auditory and visual domains.

To achieve our vision of visual support for musical practice, we need to address these challenges. 
As a first step in this direction, this paper focuses on two goals: 
First, we aim to qualitatively demonstrate the value of such feedback for musicians as a general means for a variety of practiced skills.
Second, based on what we learned, we want to provide considerations that help designers of similar applications better understand the abundant design space.
We sought to explore a broad range of applications and design alternatives to understand the problem~\cite{gleicher2023problemspace} and design spaces. 
Therefore, we conducted a design space exploration, rather than following a more traditional design study methodology~\cite{sedlmair2012design} focused on a single application goal.

We conducted our design exploration over the course of five years (\cref{sec:overview}).
Building on our own experience as musicians and visualization researchers, we created and tested early designs over multiple projects where we worked with other musicians to learn about the space of possibilities.
In a final convergence phase, we created a collection of small designs that sample interesting parts of this design space formed by various aspects of music practice, such as different instruments, data sources/formats, genres, human factors like motivation, interests, technical ability, and musical skill level, as well as visualization and interaction options.
As this space is vast, we narrowed down our scope and focused on MIDI~\cite{moog1986midi} and audio pitch, three different instruments (keyboard, drums, guitar), a subset of musical skills, and immediate visual feedback that can be used during practice sessions with minimal data and interaction. 

Our convergence phase resulted in 33 designs that we implemented as prototypes.
In this paper, we present a few of these as examples that best illustrate ways in which visualizations can make difficult-to-hear patterns easy to see (\cref{sec:design-exploration}).
The remaining are described in the supplemental material.
These example designs serve as small case studies that exemplify the range of possibilities of music practice visualizations. 
To better understand our approach, we conducted a qualitative study with both music learners and teachers (\cref{sec:expert_evaluation}).
\chgmin{%
    The results suggest that visualizations can make the unhearable visible, that this has value in music practice, and that the approach may have roles in music education.
}{%
    This study aimed to explore the broader potential of a visual approach to support music learning, as opposed to scrutinizing individual visualization prototypes or their usability.
    The results suggest that visualization can make the unhearable visible, that this has value in music practice, and may have a role in music education.
}
This paper provides lessons learned during our endeavor in the form of design considerations for designers of similar practice support applications (\cref{sec:considerations}). 
To aid reproducibility and allow musicians to benefit from our outcome, we published our source code and a web app with prototypes at \href{https://github.com/visvar/mila}{github.com/visvar/mila}.

We make the following core contributions: 
1)~a~design exploration that demonstrates a range of possible visualization designs for music practice and provides examples of how unhearable things can be made visible,
2)~evidence for their potential, collected from the use of our designs, and
3)~considerations for future research and products.

\section{Related Work}

\subsection{Music Visualization}\label{sec:relwork-musicvis}
There is a broad range of applications for music-related visualization~\cite{khulusi2020survey}. 
A large portion is focused on musicology and helping to understand the structure~\cite{miller2024melodyvis, heyen2023visual, cantareira2016moshviz}, \chgmin{}{semantic sequences~\cite{chang2023muse}}, and theory~\cite{miller2022augmenting, bergstrom2007isochords, deprisco2017understanding} of finished pieces.
Some work supports the analysis of professional performances of pieces or improvisation~\cite{snydal2005improviz}.
Others use visualization for the exploration of larger music collections~\cite{miller2022corpusvis} or personal listening behavior~\cite{baur2010streams}.
As music is inherently temporal, many challenges and approaches overlap with visualization of time-oriented data~\cite{aigner2011timevisbook}.
While there is work on visually augmenting sheet music~\cite{watanabe2003brass} and instructions to support learning, the visualization community has not addressed feedback on practice data and its inherent messiness yet. 

\subsection{Instrument Practice Support}

The research on computer-aided practice support for musical instruments has mostly been conducted in the field of human-computer interaction. 
Examples of such approaches use playing or sensor data to prevent injury~\cite{ren2019forceguitar, fender2023pressurepick} or adjust the difficulty of exercises~\cite{maitz2023neuroadaptive, karolus2018emguitar}, or gaze data to highlight difficult sections of sheet music~\cite{karolus2023eyepiano}.
The more recent FretMate~\cite{shu2025fretmate} uses ChatGPT to provide personalized assessment on pitch, rhythm, and chord fingering as well as structured exercises and adaptive motivation.
In contrast, we focus on supporting self-assessment through visualization of what is played.
Some visual feedback approaches augment sheet music~\cite{asahi2018toward:piano:support, hori2019piano:hmm, smith2008interactive, fober2007vemus} to visualize assessments through coloring, highlighting, and comparison and therefore only work for exercises where musicians follow pre-defined notes closely.
Often, feedback is only available for a narrow range of musical skills.
A timbre visualization~\cite{arai2023timtoshape} guides a musician by comparing their sound to a target on a 2D space.
\emph{Strummer}~\cite{ariga2017strummer} focuses on guitar chords and \emph{Soloist}~\cite{wang2021soloist} on guitar solos, where the user's playing is compared to audio from instructional videos.
Closer to our approach is an app~\cite{ng2015easytolearnpiano} that helps children learn to play the piano, train their ears, and compose simple melodies.
Other research has explored different modalities for visual feedback beyond screens, such as LEDs~\cite{keebler2014fretlight, marky2021letsfrets}, projectors~\cite{yamabe2011feedback}, and virtual reality headsets~\cite{heyen2022cellovis}.
Augmented instruments~\cite{deja2022surveyaugmentedpiano} were designed to show instructions~\cite{loechtefeld2011guitar, keebler2014fretlight} and feedback~\cite{marky2021letsfrets}
or directly improve the playing~\cite{xia2018shift}, for example, by correcting the pitch~\cite{pardue2019separating}.
Even though some of the above work uses visual encodings, these are mostly simple. 
As researchers with a background in visualizations and as musicians ourselves, we take a visualization design perspective~\cite{munzner2014bookvad} and broadly explore possibilities to show more detail and context.
\chg{}{As a result, the designs we created are more abstract than those from related work.}

\subsection{Commercial Products}\label{sec:commercial}
Outside of research, several commercial products, such as learning apps and games, are available.
Instrument education games~\cite{soszynski2016music} such as Rocksmith (\href{https://rocksmith.ubisoft.com/}{rocksmith.ubisoft.com}), Yousician (\href{https://yousician.com/}{yousician.com}), Synthesia (\href{https://synthesiagame.com/}{synthesiagame.com}), and PianoVision (\href{https://www.pianovision.com/}{pianovision.com}) have been shown to support learning~\cite{jenson2016explorationmusicvideogames}, but only provide immediate feedback or are limited to simple scores or charts.
Some games are limited in realism~\cite{arsenault2008guitar}, as they focus more on the experience than \chgmin{actual}{} learning~\cite{miller2009schizophonic}.
Those that use actual instruments aim more towards education~\cite{graham2018rockgod} and can shape how players reason about instrument playing~\cite{omeara2016rocksmith}.
However, the demands of gamification often lead to simple feedback without context; we seek to explore richer feedback connected to a broader set of goals.

\section{Overview}\label{sec:overview}

In this section, we provide an overview of our method.
We then explain and categorize the kinds of unhearable patterns we want to make visible.
Last, we list design requirements and discuss why we use case studies as evaluation.

\subsection{Research Through Design}

To understand the space of possibilities for how to make unhearable patterns visible, we sought to explore a broad range of applications and design alternatives. 
We approached this goal through a design space exploration rather than following a more traditional design study methodology~\cite{sedlmair2012design} focused on a single, more specific application.
Our design space exploration combines an auto-biographical~\cite{neustaedter2012autobiographical} approach using our own expertise as musicians and educators with more traditional user-experience research strategies of sampling our intended audience.
The authors are all musicians at varying levels and one author has seven years of experience as a music teacher. 

Throughout our work, we identified different kinds of patterns that are hard to hear (\cref{sec:unhearable}).
Furthermore, we collected ideas, inspiration, and feedback from different research projects in collaboration with 18 musicians (not including the 13 participants of our study, see \Cref{sec:expert_evaluation}) and from multiple student theses over the course of five years. 
Grounded in this knowledge, we chose a scope and design requirements (\cref{sec:requirements}) and conducted research through design~\cite{stappers2014researchthroughdesign} by intensively iterating on our designs using sketches, prototypes, and design critiques\chgmin{}{, which we got by presenting ideas and prototypes within our institute and by demonstrating them to guests}.
Through annotations and tables, we systematically structured our exploration and filled interesting gaps in the coverage of possibilities by creating further designs where reasonable.
We view our work as a step towards a structured and abstract design space characterization. 
Moreover, we do not consider our exploration to be comprehensive, but a first step toward a better understanding of the design space.
A more structured and abstract design space characterization is ambitious and left for future work.

\subsection{Unhearable Elements in Musical Practice}\label{sec:unhearable}


The overall experience of music depends on many elements coming together. 
A musician must learn how to control these often subtle details and how they combine to create this experience. 
However, these individual elements can be difficult to hear by themselves but \chgmin{can be still}{still} important to \chgmin{overall}{} music learning -- we term them as \emph{unhearable}.
We divide the unhearable patterns we identified into two categories based on their origin:
those stemming from limitations of auditory perception (hearing) and those caused by limited cognition and memory.

\paragraph{Limitations of Auditory Perception}
This category includes patterns that are too fine to hear or ``hidden in the noise''.
However, just because the (potentially inexperienced) musician cannot pick them out during practice or playback does not mean that an audience would not notice them, perhaps subconsciously or due to a different sound setup.
In other cases, the practicing musician might hear \textit{that} something is wrong, but not \textit{how}.
Moreover, even things that an audience cannot hear can reveal issues in technique (\cref{sec:design-timing-consistency}).
Therefore, it is challenging to balance between showing too little and too much -- to show exactly what matters.
\begin{itemize}
    \item Small deviations, such as small timing mistakes, can be visualized as clearly as needed.
    \item Different patterns overlayed in mixed signals, such as differences between drums played by hands and feet, are hard to hear since sounds overlay, but mistakes might only happen when playing with both. A visualization can show them separately.
\end{itemize}

\paragraph{Limitations of Memory and Cognition}
Under this category fall all patterns that happen over a time of more than a bar or a few seconds.
Extracting the necessary information would require summarizing/integrating over these longer time spans or comparing between time points that are farther apart.
So, while these patterns could be heard when listening multiple times or jumping around in a recording, a visualization can save effort and time by showing overviews of multiple time spans or the complete recording \textit{at once} -- which supports comparison and guides the user to what is worth listening to.

\begin{itemize}
    \item Trends, such as slowly getting faster unintentionally -- a common problem when playing without a metronome since time perception is imperfect. Comparison to a baseline helps see such trends.
    \item Variation over time, such as how varied or repetitive an improvisation chooses note pitch and duration. Visual overviews can show variation and compare multiple recordings.
    \item Distributions, such as how often each part of the instrument was used while improvising for a few minutes. Visualization can summarize long time spans in a single image and represent the playing in a familiar layout.
\end{itemize}

\subsection{Design Requirements and Scope}\label{sec:requirements}

Based on our experience from early designs, we chose a set of requirements.
These are not recommendations for instrument practice visualizations in general, but necessary for us to reduce the scope to make exploration feasible while still keeping enough breadth.

\chg{}{%
Our work focuses on helping musicians reach their goals in practice drills, which are a key part of learning important skills.
Later, musicians may intentionally ``break the rules'' to achieve musical effects.
For example, the ability to play an even timing in a scale drill may sound mechanical, but once a musician learns to control their timing, they can use this control to create intentional variation for expressiveness.  
In other words, our designs are meant to give feedback on how well the musician sticks to certain drills and expectations that they might define themselves or get assigned by a teacher.
}

\paragraph{\chg{}{MIDI instruments}}
We chose to support three different types of instruments for which we have the means to get MIDI~\cite{moog1986midi} output: digital piano keyboards, electronic drum kits, and electric guitars fitted with a special pickup.
MIDI provides discrete notes with onset, duration, pitch, loudness, and additional information, such as the drum or guitar string they were played with.
We almost exclusively use MIDI because these features are hard to reliably extract from audio.
Further instruments could be supported through future hardware or software extensions.

\paragraph{\chg{}{Immediacy}}
To facilitate testing our designs, we focus on feedback either during playing or immediately after an exercise\chgmin{is completed}{}.
We avoid scenarios that require larger amounts of data, post-processing, or user interaction for data management.

\paragraph{\chg{}{Familiarity}}
\chg{}{%
We have chosen to focus on simple, standard visualizations because we believe that these are more appropriate for our target audience who may not be familiar with, or able to understand, more complex visualizations.
Concretely, we mostly build upon pie, line, bar, and area charts, as well as encodings that are similar to notation or instruments.
Some musicians are familiar with piano rolls and waveforms, which are commonly used in music software.
We use piano rolls and simplified similar layouts, but did not include waveforms, as we strive for more abstract visualizations.
}

\paragraph{Isolation of skills}\label{sec:isolationofskills}
We chose to focus on individual skills because musicians often practice with drills that isolate elements they work on.
Therefore, most of our prototype designs each focus on one skill, rather than the interplay between them (for an example of combining two, see \Cref{sec:design-accents}). 
    In the following, we list the categories of skills we support with examples:
    1)~\textbf{Timing}: holding notes for a duration, playing along a regular sub-division of a bar (e.g., eighth notes, triplets), playing an uneven rhythm like a swing feel, playing multiple rhythms at once, keeping or changing tempo.
    2)~\textbf{Pitch}: exactly playing the intended pitch, bending (smoothly transitioning from an initial to a target pitch), vibrato (oscillating quickly around a pitch).
    3)~\textbf{Dynamics}: holding constant, in- or decreasing, accenting by playing certain notes louder (e.g., every third one in a triplet pattern).
    4)~\textbf{Improvisation}: picking notes matching the intended sound by choosing from intervals, scale degrees, or notes that fit a chord progression (improvisation skills are more vague, as there are no strict ``rules'' to follow).
    5)~\textbf{Instrument}: knowing where notes are played on the instrument, knowing which strings should be strummed and in which direction (for guitar).

\subsection{Qualitative Evaluation Through Case Studies and \chgmin{Expert Interviews}{Interviews With Music Learners and Teachers}}

Our objective is to study how visualization can be designed in different ways to reveal unhearable patterns. 
Therefore, we focus on exploring design alternatives and assess their success through direct observations. 
System-level usability and ease of learning would be important in a deployable system; however, it is not a primary assessment criterion at this point.
Instead, the criterion is whether visualization can show us patterns that are hard to hear but relevant for practice.

Therefore, we assessed our dynamic and design-focused process with anecdotal evidence and examples as case studies that we gathered along the way, a common \chgmin{evaluation}{} practice in data visualization~\cite{isenberg2013systematic}.
This evidence includes over 1500 recorded snippets of various sizes from different musicians with diverse expertise as well as synthetic data for characteristic problems.

We also conducted a more structured study with music learners and educators who were not part of the development process (\cref{sec:expert_evaluation}).
In these interviews, we evaluated
1)~whether our designs succeed at making the unhearable visible and whether this would be useful, and
2)~how our approach would benefit their musical practice or teaching.

\chgmin{
}{%
    To summarize, we ``triangulated'' the evaluation from different angles:
    Through auto-biographical design, we made sure that our collection is useful at least for a small group of musicians. 
    Additional feedback from participatory design allowed us to extend and improve this collection.
    We then evaluated whether musicians who were not part of the design process can see relevant patterns in the visualizations.
    Finally, we further interviewed the participating musicians who have teaching experience and discussed how the visualizations could be used in educational settings.
    Our case studies and evaluation with musicians demonstrate the broader potential of the visual approach, but we did not directly evaluate usability or adoption.
}

\section{Design Exploration}\label{sec:design-exploration}

In this section, we describe a selection of our designs in the form of case studies.
We chose them because they represent different examples of what can be hard to hear but easy to see and led to interesting takeaways, both in terms of visualization design and visualization-supported music practice.
\chgmin{We use inline figures without captions in this section for better integration with the text.}{}
For a video with these and more examples, see our supplemental material at \href{https://github.com/visvar/mila}{github.com/visvar/mila}.

\subsection{Note Duration}\label{sec:duration-pies}

\paragraph{Problem}
The duration of notes is a fundamental part of rhythm.
Here, we focus on the actual duration of the note -- how long it is held between onset and release -- and not the timing of and between notes.
When learning note durations, beginners need to understand whole, half, and quarter notes ( {\small \musWhole, \musHalf, \musQuarter\ }) and concepts such as dotted notes.
For example, a quarter note is half the duration of a half note; a dotted quarter note ( {\small \musQuarterDotted\ }) is 50\% longer than a quarter.
While holding a note for a certain duration is more important for key and wind instruments, guitarists need to practice stopping notes for rests (silence).

There are two hard-to-hear issues:
\textit{1) Playing the wrong duration.}
A~common issue is counting correctly for the onsets but not holding the note long enough. 
Beginners struggle with counting and playing at the same time, and dotted notes require counting in more complex patterns (1-e-and-a, 2-e-...), which needs practice.
\textit{2) Playing a roughly correct duration too short/long}.
This issue is caused by untrained auditory perception and coordination between hearing and playing.

\paragraph{Design}
Comparing a played note to correct durations requires setting the tempo, a factor that translates musical time to seconds: 120 beats per minute makes a quarter 0.5 seconds long (in common time).
Musicians often play along \chg{}{with} a metronome that makes a click/beep sound at this rate. 
As we are only interested in note duration, they can play arbitrary notes -- we only visualize the time between onset and release.

\begin{figure}[htb]%
      \centering %
      \includegraphics[width=\linewidth, alt={Four pie charts filled to a quarter, half, three quarters, and full. All have ticks in steps of a 1/8 circle, and the first has note symbols for quarter, half, 3/4 (dotted half note), and a whole note.}]{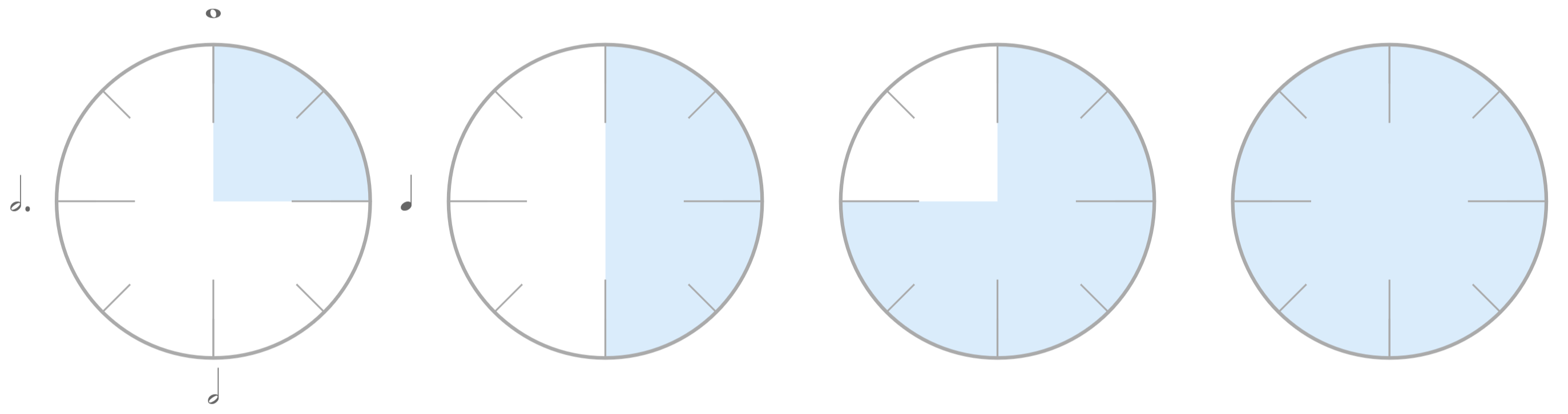}
      \caption{%
      	\chgmin{}{Note duration pie charts for perfectly played quarter, half, dotted half, and whole note.}%
      }
      \label{fig:durationpies}
    \end{figure}

A straightforward way to encode note duration as the part of a whole (note) would be a progress bar.
Instead, we chose a pie chart encoding\footnote{\chg{}{Duration: \href{https://visvar.github.io/mila/?d=duration-pies}{visvar.github.io/mila/?d=duration-pies}}}, inspired by a high school music teacher who told us that a pie metaphor is already used in educational material to explain note durations.
The visualization of each note is only shown after it is released, to provide immediate feedback while avoiding reliance on learning by eye instead of ear.
\chgmin{%
Perfectly played quarter, half, dotted half, and whole notes would look like this:
}{%
\Cref{fig:durationpies} shows how perfectly played quarter, half, dotted half, and whole notes would look like.
}

\begin{figure}[htb]%
      \centering %
      \includegraphics[width=\linewidth, alt={Multiple bars that are filled with a blue area. The first few are increasingly getting closer to a tick mark that indicates the duration of a dotted quarter note (the middle between a quarter and a half note). The last two bars are attempts at a whole note; one is not quite fully filled, and the other is overflowing, as indicated by an additional dark blue area.}]{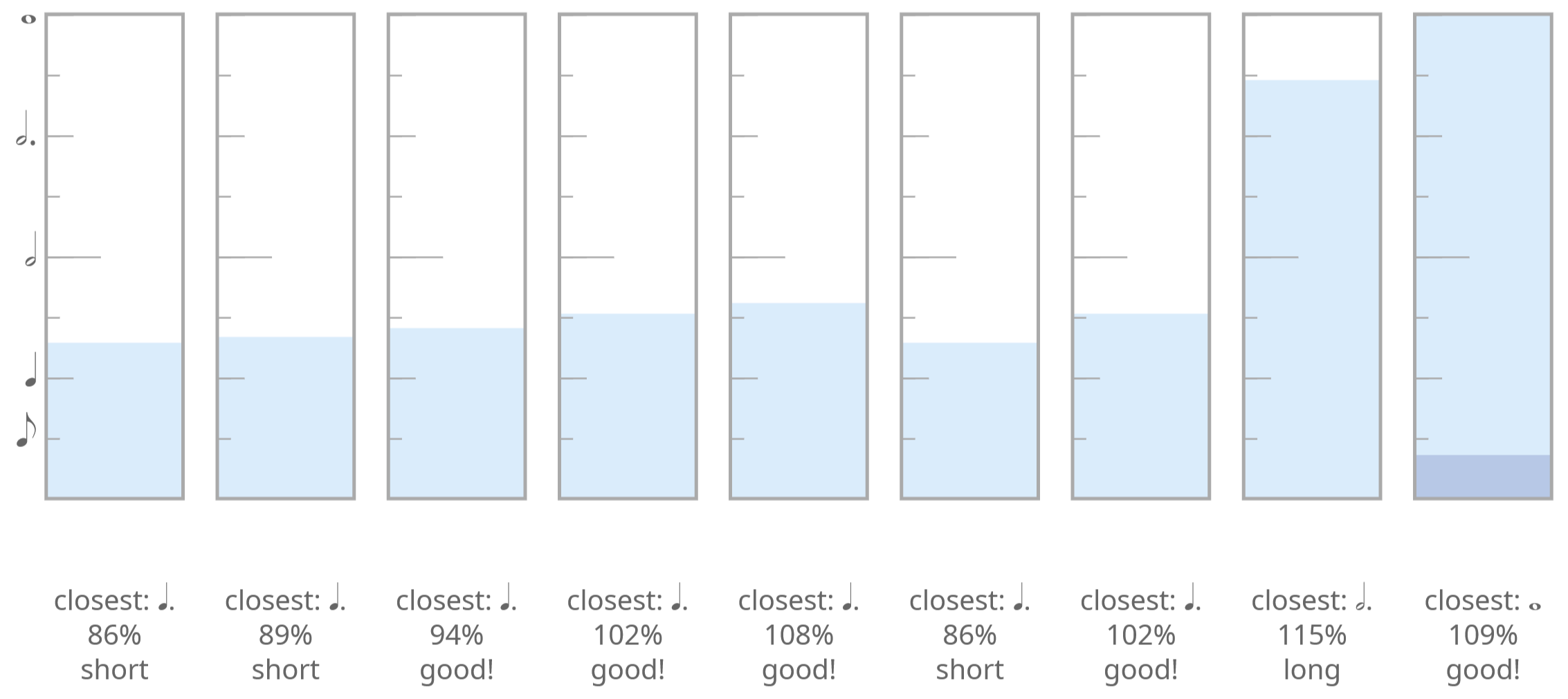}
      \caption{%
      	\chgmin{}{Note duration bar chart showing seven attempts at playing dotted quarter and two at a whole note.}%
      }
      \label{fig:durationbars}
    \end{figure}

The visualization only shows how close a note is, but does not tell how \textit{good} it is -- as a teacher would.
To explore automatic assessment, we added textual feedback with three cases: too short, too long, and good, with a threshold of 10\% of the closest correct note duration.
%
%
The pie chart encoding has limitations: 
When trying multiple times to play the same duration, comparing the angles/areas between notes is harder than \chgmin{it would be}{} with a linear encoding, which would also be more space-efficient.
Therefore, we created a variation with progress bars that uses height \& area instead of angle \& area and follows the mental model of a glass filled with water \chgmin{}{(\cref{fig:durationbars})}.

\paragraph{Examples}
What might a musician learn from the visualization?
\chgmin{The image below}{\Cref{fig:durationbars}} shows the bar encoding and the scores for a musician trying to play a dotted quarter and then a whole note.
After initial short notes follow some that fall within the threshold (labeled ``good'').
The first attempt at a whole note is just barely short enough to be closer to a dotted half and, therefore, assessed as too long (there is no knowledge of what the target duration was).
A second attempt was longer than a whole -- in this case, we fill the pie/bar again with a darker color to indicate an overflow\chgmin{:}{ (\cref{fig:durationbars}).}

\paragraph{Take-Away}
From this admittedly simple design, we can learn the following lessons:
1) Visualization can exploit (visual) mental models from educational material.
2) Simple encodings can work well -- there is no need to involve sheet music notation for learning musical concepts (besides the symbols for duration).
3) Trade-off: bars might be easier to perceive~\cite{blascheck2023parttowhole} and compare, whereas the pie encoding better fits the existing mental model (and was preferred by the musicians we asked).
4) Even simple automatic assessment is challenging: should the threshold depend on the tempo? On the target duration? On the instrument (timbre)? On the musician's skill level? After all, it depends on perception, and perception is influenced by all these factors.

\subsection{Timing Consistency}\label{sec:design-timing-consistency}

\paragraph{Problem}
Beyond duration, there is another important aspect of timing -- the onset, or start, of the note.
A musician needs to be able to control this rhythmic timing, either to follow a certain pattern (for example, when playing to a grid) or to vary it for expression.
The goal is not to achieve perfection (which could even sound robotic and unnatural, see drum humanizers), but to be able to play consistently.
\chg{}{%
Musicians typically first practice to develop control and then use it to play expressively.
}

A metric for how well a musician controls timing is how consistently they can repeat a rhythmic pattern.
While solo pianists are rather free in their variation of timing, musicians who play with others, especially drummers, invest a lot of effort into practicing precise note onsets.
However, there are two reasons why hearing alone makes it hard to analyze consistency:
1)~The musician's auditory perception might miss fine details such as slightly off-beat notes during practice that an audience would notice. 
This can happen because they are focused on reading or remembering sheet music and their movements while playing, or simply because their ears \chgmin{not}{are} not trained yet. 
While such details could be heard in a recording, it requires time and focused listening.
2)~Studying consistency over multiple repetitions would require remembering and integrating over a time span of a few bars to multiple minutes -- this is even hard when listening to a recording.

\begin{figure}[htb]%
      \centering %
      \includegraphics[width=\linewidth, alt={A screenshot of  audio software showing a recording as waveform. Next to it is the same waveform cut in the middle and moved to form two rows. Brighter colors highlight the beats to show how the two rows correspond to the same beats, only in two repetitions.}]{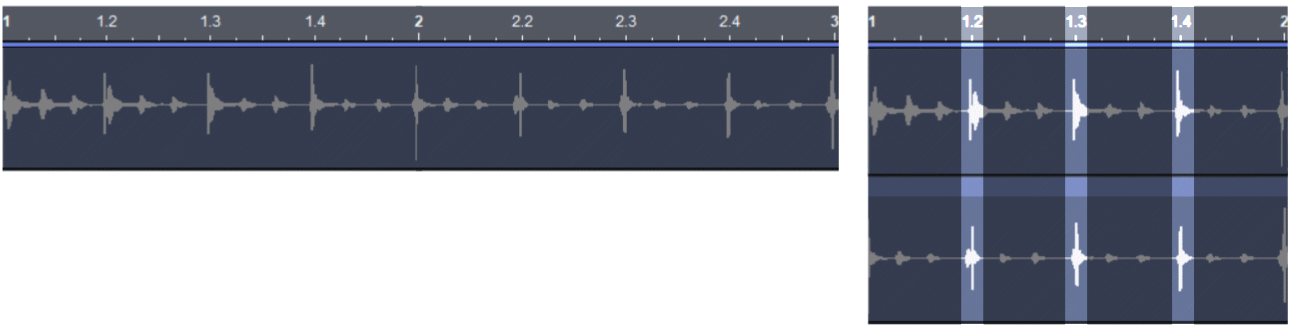}
      \caption{%
      	\chgmin{}{Aligned stacking of repetitions.}%
      }
      \label{fig:stackedrepetitions}
    \end{figure}

\paragraph{Design}
We focus on a common kind of exercise for rhythm practice, where the same short pattern is played repeatedly to a metronome.
An existing, ubiquitous visualization we could use is a waveform -- an area chart that encodes the loudness of audio over time from left to right.
However, comparing between a few repetitions would already require looking back and forth and potentially panning or zooming.
Exploiting the repetitive, cyclic time, we can keep displaying time from left to right and use a new row for each repetition, which vertically aligns the same beats \chgmin{(highlighted):}{(highlighted in \cref{fig:stackedrepetitions}).}

\begin{figure}[htb]%
      \centering %
     \includegraphics[width=\linewidth, alt={Four rows with ticks, with each row, the ticks are increasingly left. Behind each column of ticks is a light gray rectangle. Except for the first row, all ticks fall within these rectangles.}]{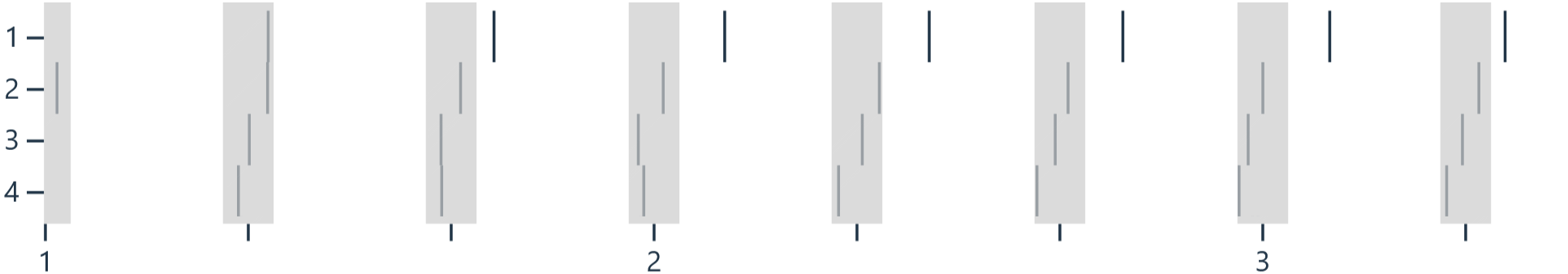}
      \caption{%
      	\chgmin{}{Tick encoding with light gray tolerance areas around the target times.}%
      }
      \label{fig:timingticks}
    \end{figure}

\chgmin{The above image}{\Cref{fig:stackedrepetitions}} shows how audio waveform would look in such a layout -- alternatively, one could use a MIDI piano roll.
As we are only interested in the note onset (and not duration, loudness, timbre), we chose a simpler encoding with ticks that is easier and faster to read\footnote{\chg{}{Timing: \href{https://visvar.github.io/mila/?d=sub-division-linear}{visvar.github.io/mila/?d=sub-division-linear}}}.
For a reference of what is good enough, we added gray vertical bars that indicate (user-definable) tolerance zones within which notes should lie.
\chgmin{In this example,}{In the example in \Cref{fig:timingticks},} the first repetition was worse than the rest, and in the following ones, the musician played notes increasingly early.

\begin{figure}[htb]%
      \centering %
      \includegraphics[width=0.8\linewidth, alt={Four different encodings that are aligned. The first shows black ticks that overplot densely. The second shows the same ticks with semitransparency, where the density is visibly higher closer to the beat times. A histogram and area chart show the distribution more clearly. Below are the ticks and gray rectangles already used in the above images. The note ticks do not all fall within the rectangles but get closer after the first few repetitions.}]{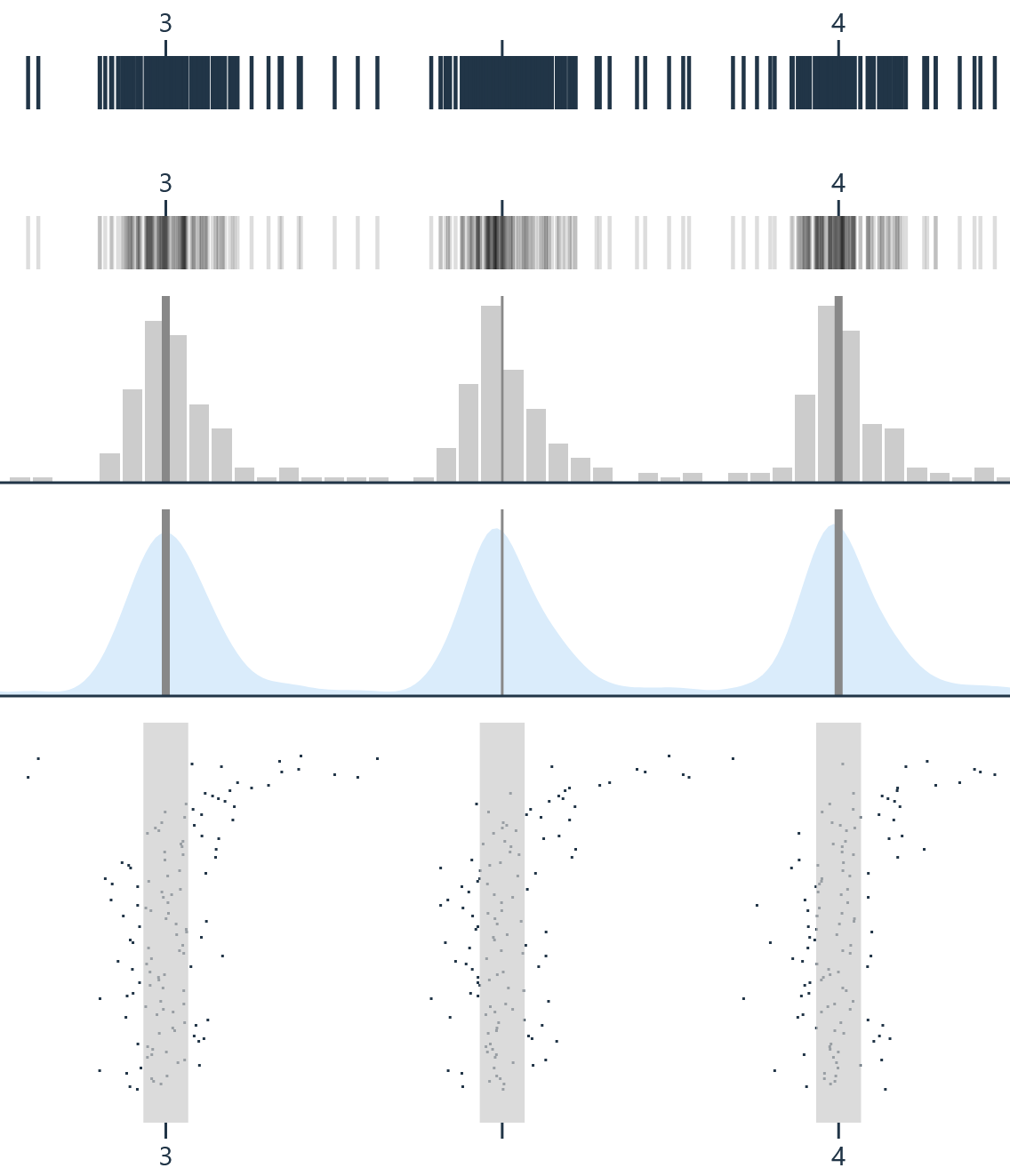}
      \caption{%
      	\chgmin{}{Timing consistency example with many repetitions.}%
      }
      \label{fig:timingmanyrep}
    \end{figure}

As with note duration before, we explored a simple automatic assessment score.
We chose the percentage of notes that lie within the tolerance because it is easy to understand \textit{and visualize} (gray bars), unlike a more complex score like Euclidean distance.
%
With many repetitions and therefore many rows, it becomes harder to see issues with consistency between beats.
To explore different ways to provide a single overview of all repetitions, we tested \chgmin{the following}{} aggregations with increasingly less visible detail \chg{(see figures below in \textit{Examples})}{(\cref{fig:timingmanyrep})}:
1)~Ticks over-plotted in a single row, without and with semi-transparency, are simple and space-efficient but become cluttered for many repetitions.
2)~A histogram can provide musically meaningful bins, such as ``a 32$^{nd}$ note off''.
3)~A~density estimation area chart (using KDE\chgmin{~\cite{rosenblatt1956kde, heer2021fastkde}}{~\cite{rosenblatt1956kde}}) is less busy than a histogram.
A drawback of all these aggregations is that they cannot reveal changes \chgmin{over time }{}across repetitions.
They are not a replacement but a supplement to the tick rows by providing an overview first~\cite{shneiderman2003theeyeshaveitmantra}.

\begin{figure}[htb]%
      \centering %
       \includegraphics[width=\linewidth, alt={The image consists of two parts: the linear and the circular version. Two blue rectangles mark the left- and right-most part of the linear version to show that notes around beat 1 get split up. In the circular version, all these notes are contained in a wedge that is indicated with an orange mark.}]{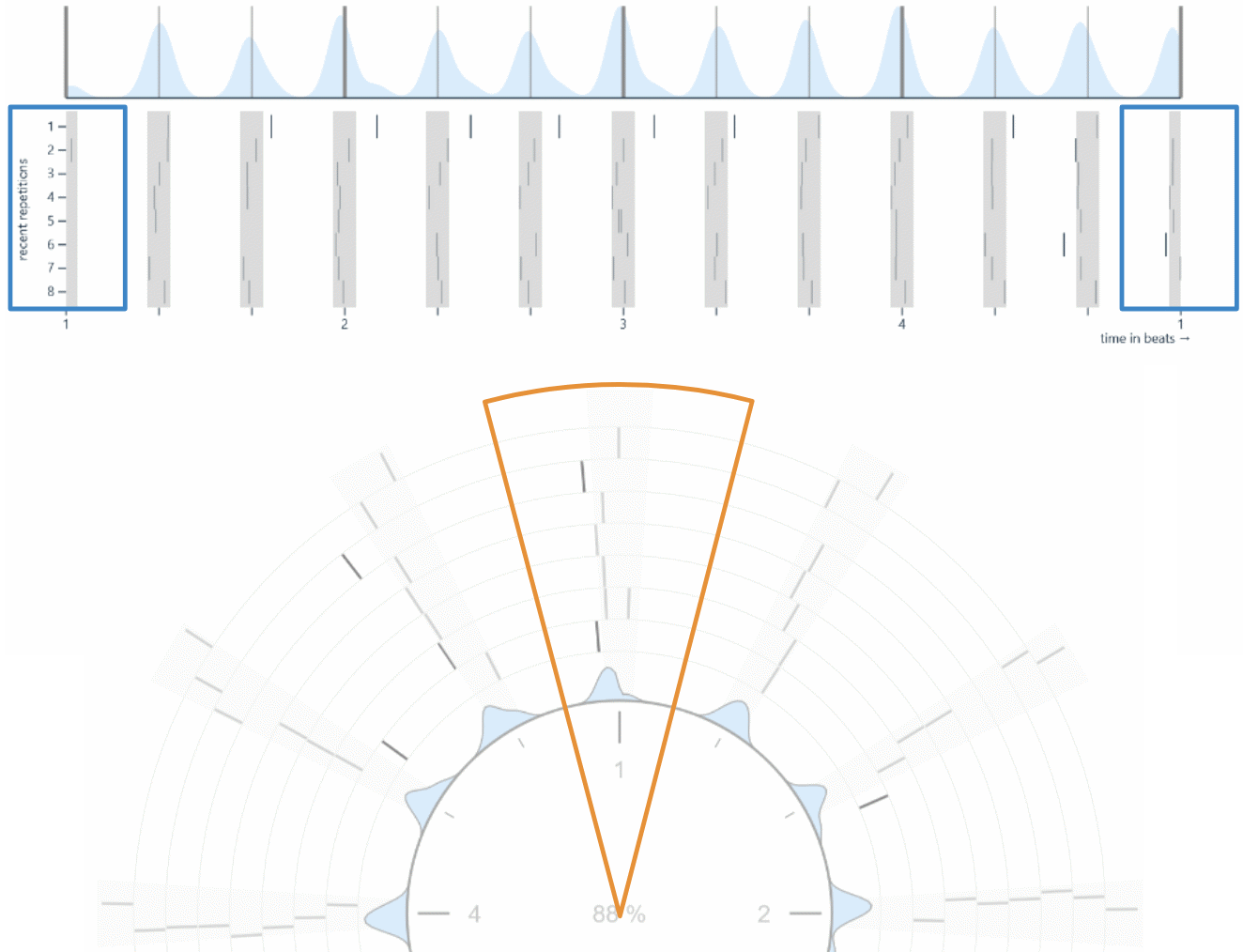}
      \caption{%
      	\chgmin{}{The circular layout avoids jumps at beat 1 but stretches time more on the outside.}%
      }
      \label{fig:subdivlinearcircular}
    \end{figure}

\chgmin{In the above case study with note duration, we compared circular and linear time encoding.}{}
We created a circular variant\footnote{\chg{}{Timing (circular): \href{https://visvar.github.io/mila/?d=sub-division-circular}{visvar.github.io/mila/?d=sub-division-circular}}} of tick rows and aggregations to address a limitation of the row layout: notes around beat 1 are drawn on the very left of a row if they are late and on the opposite side if they are early, making them hard to analyze together.
A circular layout avoids this jump but makes comparison between repetitions harder because time gets stretched more toward the outside\chgmin{.
The image below shows both variants with the same data:
}{
 (\cref{fig:subdivlinearcircular}).
}

We created further variants that show only an aggregation, but allow to compare or separately analyze multiple aspects:
1) A drummer might have different issues on different drums, so we show each kind (kick, snare, hi-hat, toms, cymbals) in a separate row\footnote{\chg{}{Timing (drums): \href{https://visvar.github.io/mila/?d=sub-division-drums}{visvar.github.io/mila/?d=sub-division-drums}}}.
2) When playing multiple rhythms at once, for example, eighth notes versus triplets with different drums or left and right hands on a piano, we display each in its own row\footnote{\chg{}{Timing (two rhythms): \href{https://visvar.github.io/mila/?d=two-handed-timing}{visvar.github.io/mila/?d=two-handed-timing}}}.
3) By comparing multiple tries of the same rhythm, which can be days or weeks apart, \chgmin{in a visualization with one row for each,}{} a musician can track progress and spot remaining issues\footnote{\chg{}{Timing (history): \href{https://visvar.github.io/mila/?d=sub-division-history}{visvar.github.io/mila/?d=sub-division-history}}}.

\paragraph{Examples}\label{sec:design-timing-consistency-examples}
\chgmin{%
    This example with over 120 repetitions shows a warm-up period, where notes were played decreasingly late for the first about 30 repetitions.
}{An example with over 120 repetitions shows a warm-up period, where notes were played decreasingly late for the first about 30 repetitions (\cref{fig:timingmanyrep}).} 
The aggregated ticks on top struggle to convey the distributions\chgmin{ around beats}{}, the histogram reveals slightly different distributions, and the density chart looks similar for all beats.
Differences \emph{between} repetitions like this warm-up are only visible in the tick rows.

\begin{figure}[htb]%
      \centering %
      \includegraphics[width=0.9\linewidth, alt={The ticks and rectangles already used in above figures showing four rows. A blue rectangle marks two columns where the first two repetitions (technique 1) produced double notes visible as additional ticks. An orange rectangle marks two other columns where notes in the last two repetitions (technique 2) were quieter, as indicated by thinner ticks.}]{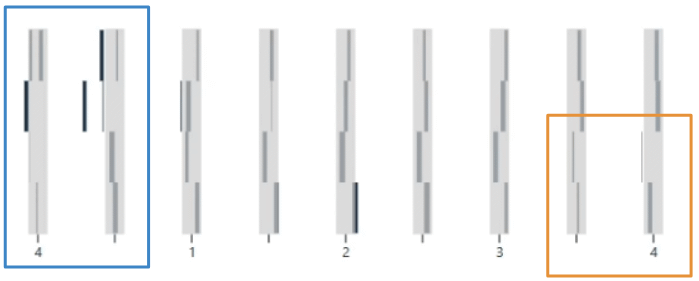}
      \caption{%
      	\chgmin{}{Double notes (blue) and loudness difference (orange).}%
      }
      \label{fig:timingdoubleandloudness}
    \end{figure}

In the following example we show an issue that might be legitimately unhearable by an audience\chgmin{}{ (\cref{fig:timingdoubleandloudness})}.
A guitarist played the same solo with two different fingering techniques (upper two rows, versus lower two rows).
While one technique led to more extra notes (blue mark), a sign of pressing strings too hard before picking, the other produced more quiet notes as indicated by thinner ticks (orange mark).

\begin{figure}[htb]%
      \centering %
      \includegraphics[width=\linewidth, alt={Multiple rows of density area charts. Going down, the distribution of onsets becomes clearer as peaks are closer to the timing grid lines and narrower.}]{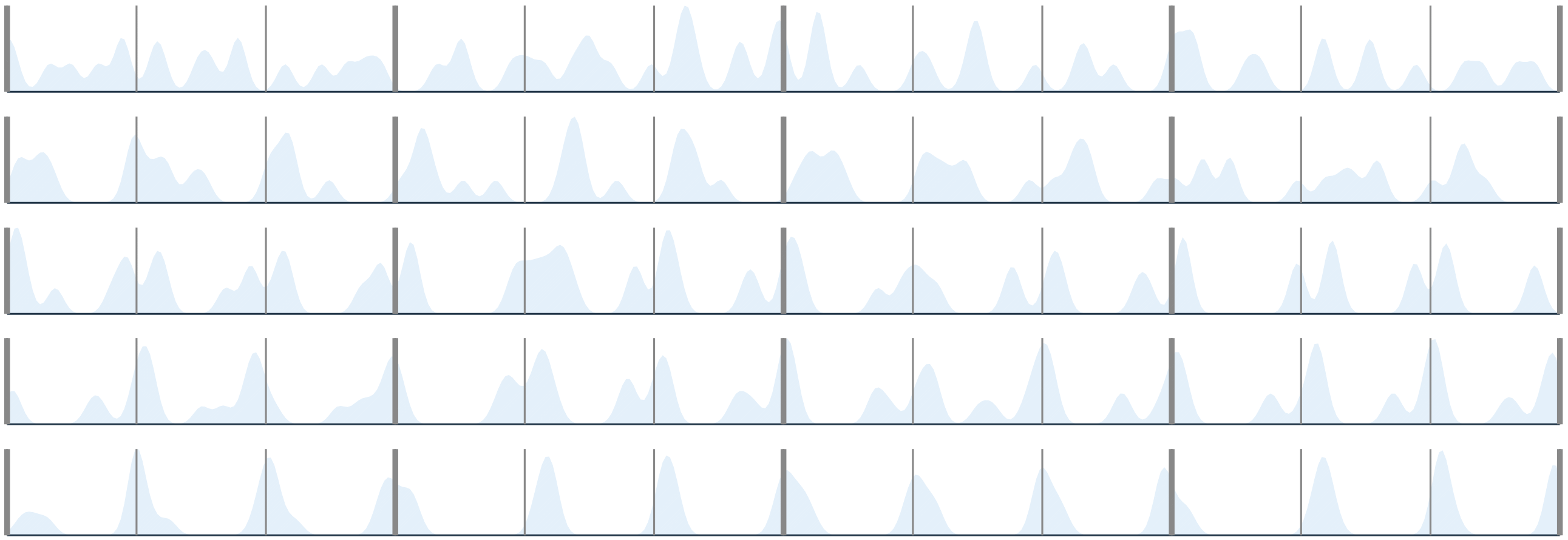}
      \caption{%
      	\chgmin{}{Comparison of five takes (top to bottom). The timing becomes more precise with each take.}%
      }
      \label{fig:timingmultipletakes}
    \end{figure}

\chgmin{The image below}{\Cref{fig:timingmultipletakes}} shows multiple takes of the same triplet rhythm.
We can see that the peaks in the lower rows (more recent takes) become narrower and closer to the grid lines, meaning that precision and accuracy improved.
Some peaks are still right of the grid lines and indicate playing slightly late.

\paragraph{Take-Away}
1) Repetition is common during practice and can be exploited for summaries and juxtaposed comparison, which is not possible with audio -- even listening to only two patterns at the same time is difficult.
2) Aggregation helps provide quick overviews and facilitates comparison between drums, hands, and voices. 
However, aggregation is not always beneficial, as patterns across repetitions like a warm-up would not be discernible.

\subsection{Accents}\label{sec:design-accents}

\paragraph{Problem}
In our next case study, we look at practicing two things together: timing and dynamics (loudness).
One special case of this fused practice is accents, notes that are played louder than the rest to reinforce a rhythmic pattern.
For example, when playing triplets, one often accents the first of each group of three notes (\textbf{one}-trip-let, \textbf{one}-trip-let,~...).
The difficulty in accenting comes from focusing on two things at once, especially for beginners.
Assessing how well and consistently one plays then also requires listening to both rhythm and dynamics at the same time (while also playing) and integrating what is heard over time.
This is again difficult for the perceptual and cognitive reasons we encountered in the above case studies, but now doubly so.
Concretely, there are the following patterns that are hard to hear:
1)~Do I accent the right notes? 
2) Do I play with consistent loudness? 
3) Do I play the correct rhythm?
4) Do I make mistakes when switching between rhythms (such as eighths and triplets) or changing the accent pattern (\textbf{one}-trip-let, two-\textbf{trip}-let, ...)?

\begin{figure}[htb]%
      \centering %
      \includegraphics[width=\linewidth, alt={Two rows of note symbols with different sizes, one large followed by three small, one large, one small, and repeating. In the first row, there are only two different sizes; in the second row, the smaller notes vary in size.}]{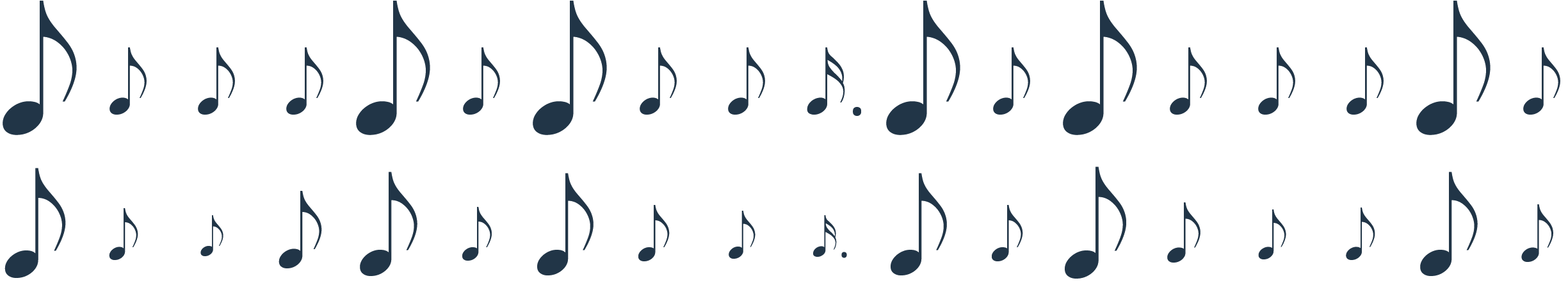}
      \caption{%
      	\chgmin{}{Top: quantized size. Bottom: continuous size.}%
      }
      \label{fig:accentsalternating}
    \end{figure}

\paragraph{Design}
As with the last case study, we assume playing to a metronome.
This time, we do not look at onsets or durations, but the time between onsets, termed inter-onset intervals or IOIs (when playing legato, durations and IOIs are almost equal, but they are different concepts).
%
We explored three different encodings \chgmin{(see figures below)}{(\cref{fig:accentsdifferentencodings})} for this data\footnote{\chg{}{Accents: \href{https://visvar.github.io/mila/?d=accents}{visvar.github.io/mila/?d=accents}}}:
1)~the ticks we found to work well before that encode onset by $x$-position and loudness through thickness,
2) two aligned bar charts that show the IOI and loudness for each note and optimize for perception of exact values, and
3) note symbols that encode IOIs similar to sheet music and map loudness to size.
We chose note symbols because they are familiar to musicians and because they quantize IOIs, which makes the visualization clearer.
This encoding also has limitations.
First, it does not scale as well as bars for many notes.
Second, compared to position (ticks) and length (bar chart), the size of symbols is harder to perceive.
However, in our case of playing accents, reading exact values is \chgmin{not important, but rather}{less important than} the comparison between different notes.
We added a variation that simplifies the note size into large and small for accents and other notes, making it easier to detect wrong accents but impossible to see loudness consistency.
\chgmin{The image below}{\Cref{fig:accentsalternating}} shows a comparison of this simplification \chgmin{(top) and linear size mapping (bottom)}{and continuous mapping}.
Here, an alternating pattern was played, where either the first or the second note of three was accented (\textbf{one}-trip-let, two-\textbf{trip}-let).

\begin{figure}[htb]%
      \centering %
      \includegraphics[width=\linewidth, alt={Note symbols for different note values are positioned from left to right based on how many beats they are long. Brackets show which ranges beat values are quantized to an eighth and to a quarter note, with the range for a quarter being twice as large.}]{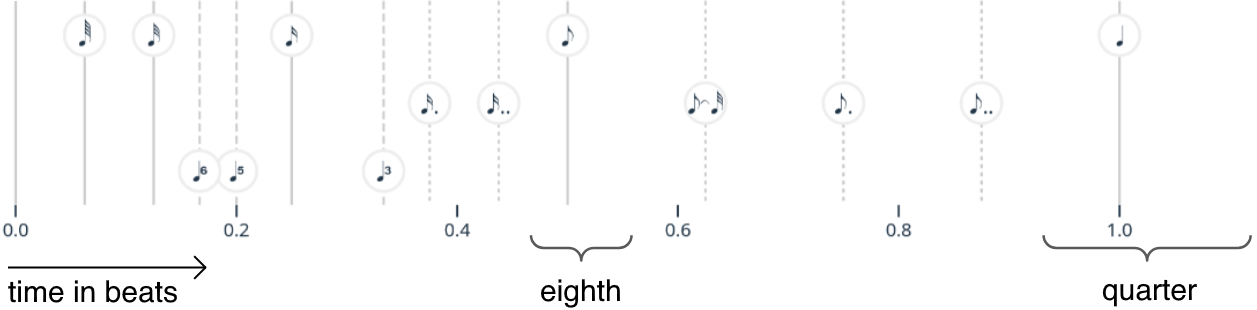}
      \caption{%
      	\chgmin{}{Representable durations.}%
      }
      \label{fig:accentsrepresentable}
    \end{figure}

\chgmin{Third,}{A third limitation is that} the IOI values that can be represented are not uniformly distributed, which makes the quantization error different for notes that are too short versus too long. 
For example, there is more leeway for quarters than eighths. 
More values are representable through modifiers, but these decrease readability. Dots add 50\%, double dots  75\%, and ties combine durations like {\small \musQuarter$\smallfrown$\musSixteenth\ } $=\frac{1}{4}+\frac{1}{16}$.
\chgmin{The image below}{\Cref{fig:accentsrepresentable}} shows note durations on a linear axis from left to right; the three rows contain simple, dotted and double-dotted, and tuplet durations.
Brackets indicate the time ranges that would be quantized to eighth and quarter notes (all durations that have these as nearest neighbors).

\paragraph{Examples}

In \chgmin{our first example below}{\Cref{fig:accentsdifferentencodings}}, we can see that the \chgmin{correct notes were accented, that is, the first of each four}{the first of each four notes has been accented (as intended)}.
On the other hand, the loudness of non-accent notes is inconsistent, as the notes following the first two accents are louder than others.
From the dotted notes, we can see that a few notes were slightly late.
For comparison, we include separate bar charts for loudness and IOI as well as ticks with loudness encoded by width (\cref{sec:design-timing-consistency}) as alternative encodings.

    \begin{figure}[htb]%
      \centering %
  \includegraphics[width=0.9\linewidth, alt={Three visualizations from top to bottom, aligned to show the same notes from left to right. The first uses note symbols and size; some notes are dotted (too long), and some non-accent notes are larger (louder) than others. Two bar charts show loudness and IOI and while differences in loudness are easy to see, differences in IOIs are less clear. At the bottom, ticks are used as in previous images, the gaps between them are slightly irregular.}]{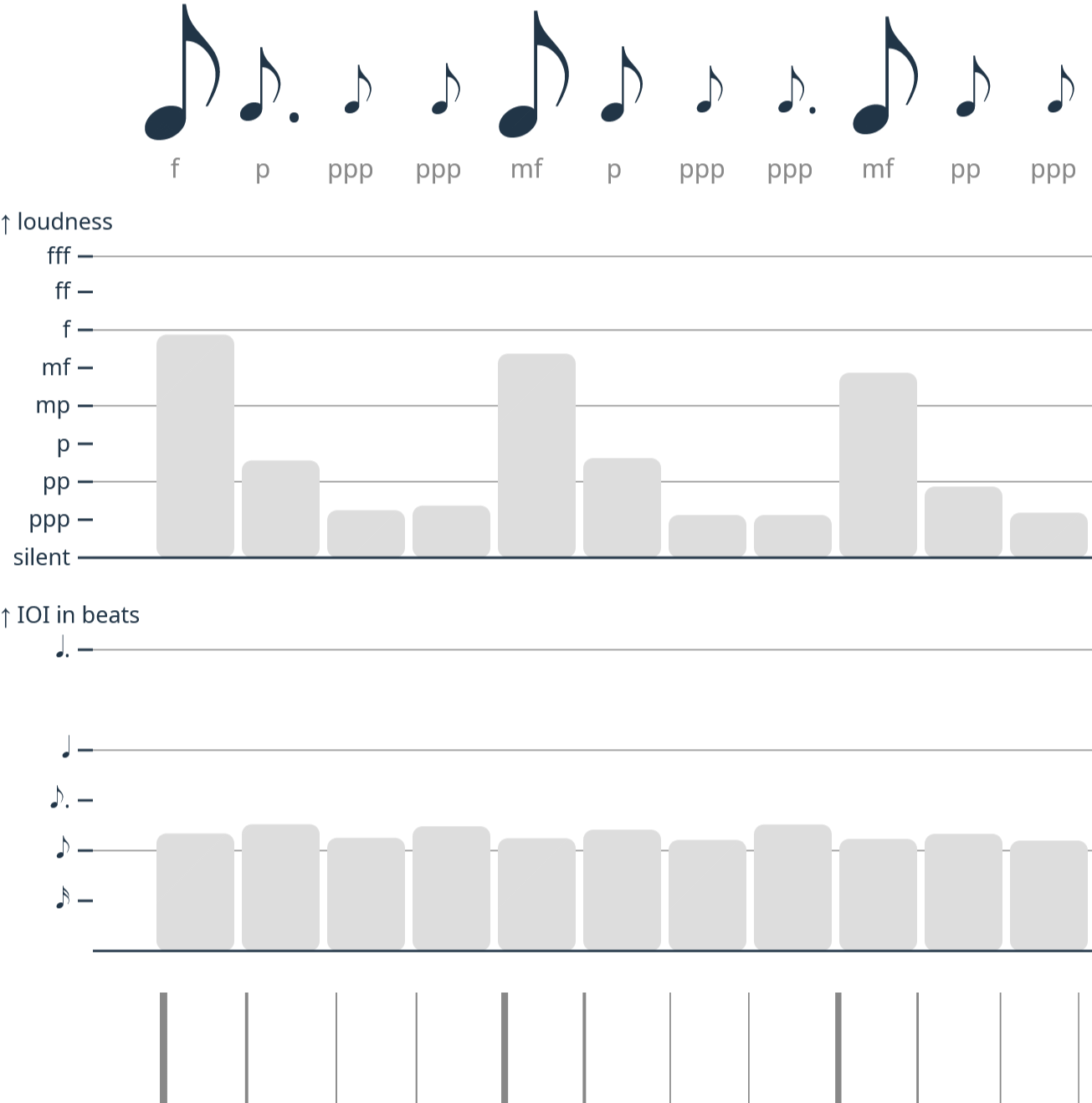}
      \caption{%
      	\chgmin{}{Comparison of different encodings.}%
      }
      \label{fig:accentsdifferentencodings}
    \end{figure}

A more difficult exercise is switching between triplets and eighths\chgmin{}{ (\cref{fig:accentstripletseights})}.
\chgmin{Below, the}{The} blue marks highlight a common mistake of playing dotted sixteenths ({\small \musSixteenthDotted}) instead of triplets ({\small \musQuarter\ $^3$}).
While the symbols might amplify the error the musician made, it could easily be missed when looking at the bar or tick encoding -- the symbols' quantization makes the deviation explicit (only the bars for IOIs shown):

    \begin{figure}[htb]%
      \centering %
    \includegraphics[width=\linewidth, alt={The image is similar to the one above, but shows a pattern where sixteenth notes show up instead of triplets that are indicated by a note with the number `3'.}]{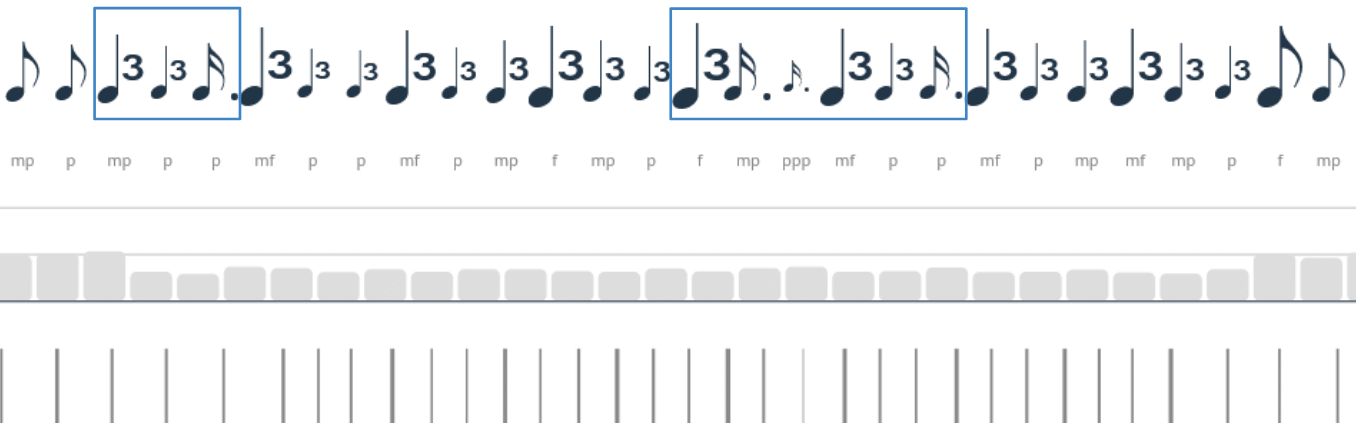}
      \caption{%
      	\chgmin{}{Example with triplets and eighths.}%
      }
      \label{fig:accentstripletseights}
    \end{figure}

\paragraph{Take Away}
1)~Visualization can support practicing two aspects at once, either with a combined (symbols, ticks) or separate encoding (bars).
2)~The note symbols exemplify the use of familiar visuals (and mental models) and adapting them:
instead of using sheet music notation as a whole, we only take a part of it and modify it through size.
3)~Quantization to discrete values, in this case IOIs, can help make potential issues more explicit and hide details like non-relevant imperfections -- but may also amplify mistakes.
While we only used it in the symbols, a quantized bar chart could bring similar benefits while scaling better with note count.

\subsection{Improvisation to Chord Progressions}\label{sec:design-chordprog}

\paragraph{Problem}
Solo improvisation, such as in jazz or blues music, requires a musician to spontaneously create music that is creatively expressive and fits within the context of the song.  
Musicians learning improvisation often begin with exercises where they try to play ``correct'' notes that fit the changing song structure, for example, by choosing notes from a scale that is appropriate for the harmony of each bar of the music. 
Within these exercises, learners often try to avoid repetitive patterns.
While good improvisation may not follow these ``rules'' (interesting solos often intentionally use ``wrong'' notes to create tension and repetitive patterns to hook the listener), novices practice these skills to develop their technique for creating solos and their ear for hearing the effectiveness of their playing.
However, even these basic exercises are difficult for a music learner to assess.   
Identifying individual notes over changing progressions or determining when a pattern is repetitive requires hearing subtlety. 
A learner needs to reflect on an entire exercise or song to appreciate their success at staying within the song’s structure with sufficient variety. 
Quantifying progress is unhearable, especially for novices. 
Indeed, part of these exercises is for the learner to develop their ear to hear the effects.

A common exercise is choosing notes from a scale that is appropriate for the harmony of each bar of the music.
To support this exercise, we have designed visualizations that show what a musician has played, relative to an underlying chord progression.
Because the tool knows the most appropriate scale for each measure, the notes played can be displayed relative to this scale, allowing the viewer to see, for example, how often they use notes that are not in the scale or overuse certain elements of the scale.
We note that this way of choosing notes is a practice drill, not 
a recipe for good improvisation.
Moreover, such simple metrics cannot tell the musician whether the improvisation sounds good or not -- only how well they did the drill.

\paragraph{Design}
After exploring different ways to use color to encode music theory, such as \chg{intervals and scale degrees within visualization (see supplemental material)}{intervals\footnote{Intervals: \href{https://visvar.github.io/mila/?d=improvisation-intervals}{visvar.github.io/mila/?d=improvisation-intervals}} and scale degrees\footnote{Scale degrees: \href{https://visvar.github.io/mila/?d=improvisation-scale-degree-colors}{visvar.github.io/mila/?d=improvisation-scale-degree-colors}} within visualization}, we chose a simplified color scheme that indicates how each played note related to the chord progression and scale.
Inspired by a musician who practices improvising with only the notes of the current chord in a backing track, we mark notes green that fit this criterion.
Those that do not, but still fit the musical scale, are colored orange and the rest are gray.

\begin{figure}[htb]%
      \centering %
      \includegraphics[width=\linewidth, alt={A piano roll with twelve rows for C, C-sharp, ... from bottom to top and time from left to right. Rectangles for notes are colored orange and green.}]{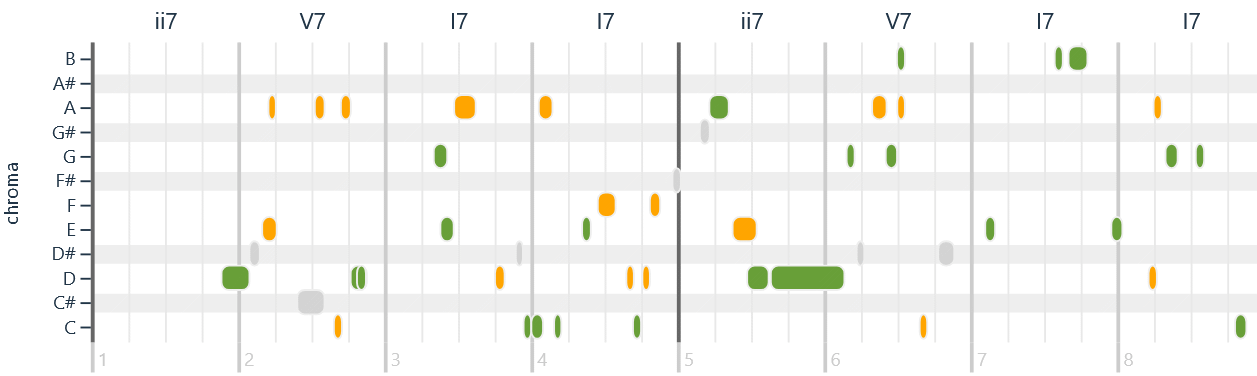}
      \caption{%
      	\chgmin{}{Piano roll with colors based in scale and chord fit.}%
      }
      \label{fig:chordprogpianoroll}
    \end{figure}

We then explored different layouts\footnote{\chg{}{Chord progr.: \href{https://visvar.github.io/mila/?d=improvisation-chord-progression}{visvar.github.io/mila/?d=improvisation-chord-progression}}}, starting with a piano roll that we extended with annotations of the current chord\chgmin{}{ (\cref{fig:chordprogpianoroll})}.
It maps chroma (note pitch without octave) to the $y$ axis, time linearly to the $x$ axis, and each note to a rectangle according to its chroma, onset, and release.

\begin{figure}[htb]%
      \centering %
      \includegraphics[width=\linewidth, alt={A visualization similar to the one above, but notes are not positioned by time and instead stacked to one stack per chroma and bar.}]{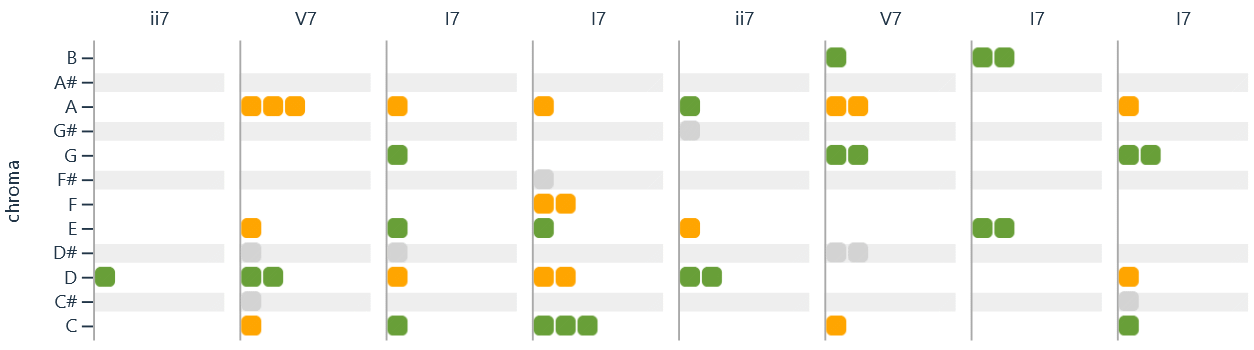}
      \caption{%
      	\chgmin{}{Waffle chart with the same data as in \Cref{fig:chordprogpianoroll}}.%
      }
      \label{fig:chordprognotecounts}
    \end{figure}

Piano rolls show the full detail but do not scale well visually for many notes or longer improvisations.
Therefore, we simplified the $x$ axis by binning the notes of each bar and showing only their count per chroma as a waffle chart\chgmin{:}{ (\cref{fig:chordprognotecounts}).}

\begin{figure}[htb]%
      \centering %
      \includegraphics[width=\linewidth, alt={Gray boxes from left to right, each containing four rows with stacked rounded rectangles that are colored green, orange, or gray. Below is the same layout with different colors. A few blue, yellow, and red dots (whole and half notes) and many blue and green dots (eighths and sixteenths).}]{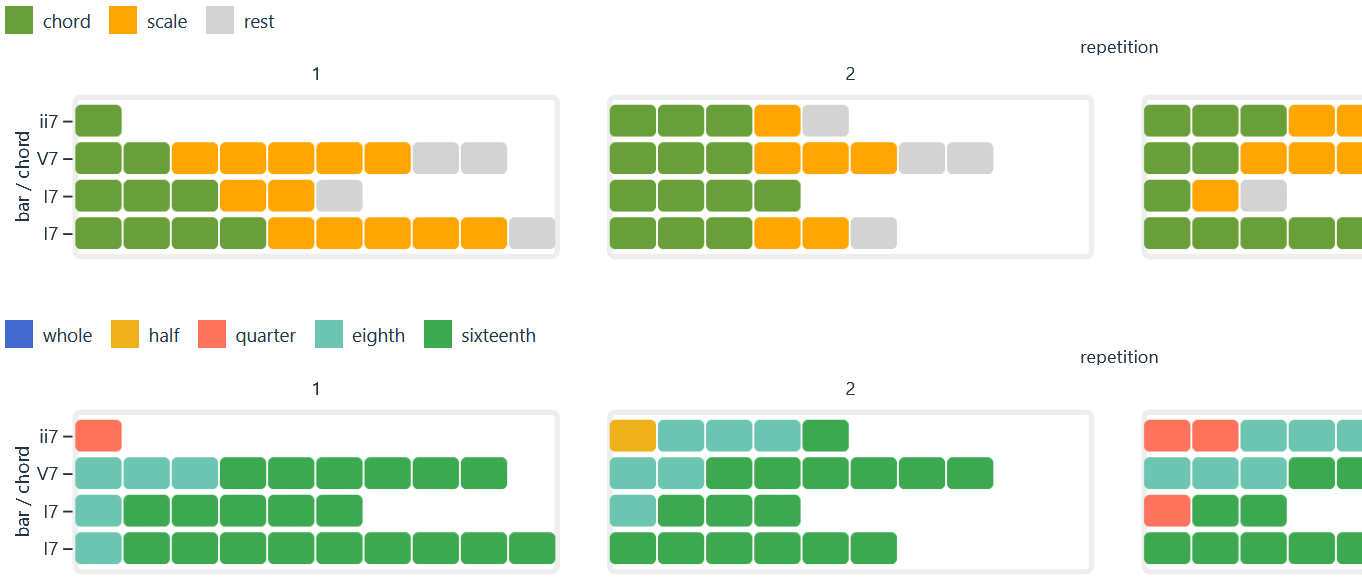}
      \caption{%
      	\chgmin{}{Layout with one facet per repetition of the progression.}%
      }
      \label{fig:chordprogfacets}
    \end{figure}

To make the layout even more compact, we \chgmin{went further and}{} added a two-dimensional binning, where each vertical bin represents a bar of the chord progression and each horizontal bin a repetition of the whole progression \chgmin{}{(\cref{fig:chordprogfacets})}.
This layout allows comparing between chords within a repetition and also between the same chord in different repetitions.
While scaling better, we lose information on the exact note chroma and only see whether they fit the chord, scale, or neither through their colors.
In contrast to the piano roll, we cannot see rhythmic information except how many notes were played.
We address this by adding a second color scheme for note duration (\chgmin{}{\cref{fig:chordprogfacets}}~bottom).

\begin{figure}[htb]%
      \centering %
      \includegraphics[width=0.6\linewidth, alt={Two rows of gray boxes filled with rectangles. The one in the top row are colored all green, those in the bottom row all blue with one red.}]{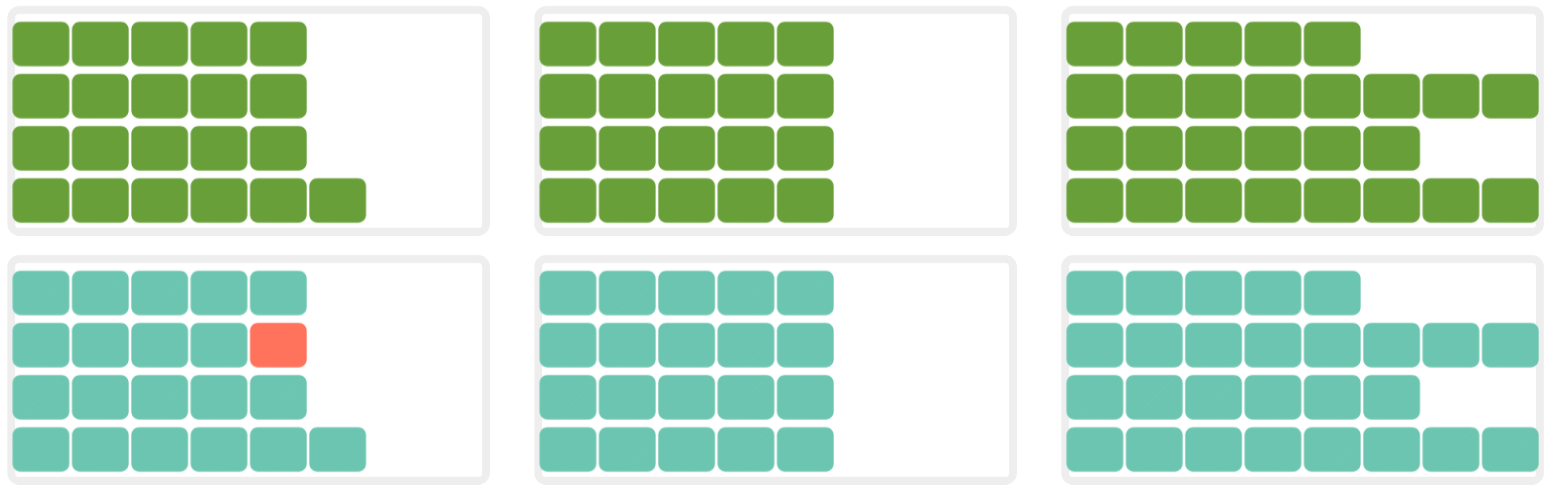}
      \caption{%
      	\chgmin{}{Improvisation with only chord notes and mostly eighths.}%
      }
      \label{fig:chordprogonlyeighths}
    \end{figure}

\paragraph{Examples}
\chgmin{The image below}{\Cref{fig:chordprogonlyeighths}} shows an excerpt from an improvisation that exclusively used green notes, that is, notes that occur in the current chord of the backing track.
When this is done as an exercise, the visualization allows quickly verifying whether one made mistakes.
The improvisation also used mostly eighth notes and would most probably sound better with some variation in rhythm.

\begin{figure}[htb]%
      \centering %
      \includegraphics[width=0.8\linewidth, alt={Three gray boxes with colored rectangles inside. The first and last have more red and blue than the one in the middle.}]{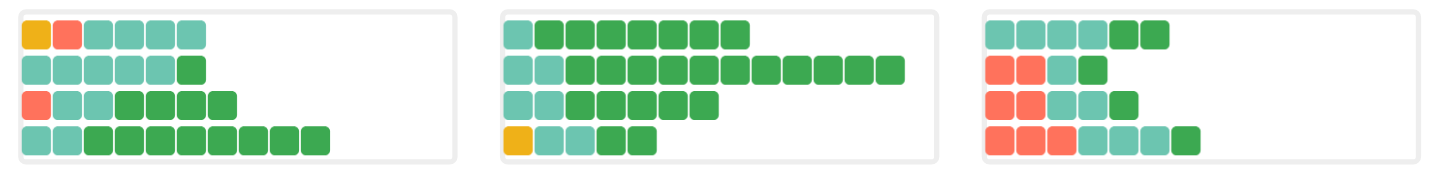}
      \caption{%
      	\chgmin{}{Switching between slow and fast.}%
      }
      \label{fig:chordprogslowfast}
    \end{figure}

In \chg{this}{another} improvisation, we see an example of switching from slower to faster and back to slower playing \chgmin{}{(\cref{fig:chordprogslowfast})}.
The first and third repetitions contain more quarter and eighth notes (red, blue), and the second more sixteenths (green).

\paragraph{Take-Away}
1) Piano rolls allow us to see exact note durations and chroma, but they become unreadable when showing more than a few bars.
A (two-dimensional) binning layout allows comparing repetitions of chord patterns and visually scales better than a linear time encoding but loses details.
2) The noise, such as short and quiet extra notes, that some MIDI instruments produce, affects piano rolls less than our other encodings, where each note gets the same visual weight no matter how long or loud.
3) One way to combine both visualizations could be multiple views with brushing and linking to combine the quick overview of the binned layout with the details of the piano roll on demand.
\chg{4)~Visualization can support ill-defined problems through comparison between what sounds good and what does not.}{}

\subsection{Movement on the Instrument During Improvisation}

\begin{figure}[htb]%
      \centering %
       \includegraphics[width=\linewidth, alt={A photo of a guitar's fretboard with strings visible as horizontal lines and frets as vertical lines. Gray circles with the letter A are placed where the note A occurs.}]{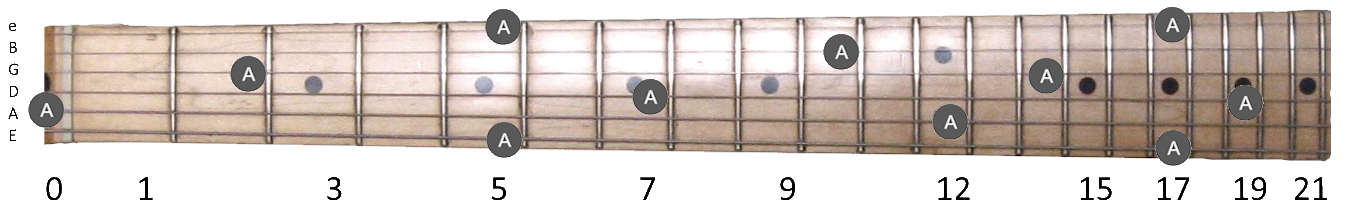}
      \caption{%
      	\chgmin{}{A guitar's fretboard with marks for the note A.}%
      }
      \label{fig:fretboardlayout}
    \end{figure}

\paragraph{Problem}
In this case study, we focus on the guitar, which has a two-dimensional layout where the same note occurs in multiple places.
\chgmin{This image}{\Cref{fig:fretboardlayout}} shows how the guitar's fretboard is commonly displayed, with 21--24 frets and usually six strings.
All positions of the note A are marked; some are identical, some in different octaves.

The way in which a guitarist moves around the fretboard during an improvisation can affect how it sounds:
When they are stuck in one region, which beginners often are, they will use only a small range of notes.
Only moving horizontally or vertically will sound less varied than mixing both.
To summarize, the goal is to analyze how the hand varies between positions and movement strategies.
By listening, even with also watching, it can be hard to get an overview of this movement for improvisations longer than a few bars.

\begin{figure}[htb]%
      \centering %
    \includegraphics[width=\linewidth, alt={A scatterplot with dots of different sizes and a regular grid in the background. The axes and grid resemble the guitar fretboard in the image above. The dots form a broad diagonal pattern from botten left to top right with colors fading from  teal to purple.}]{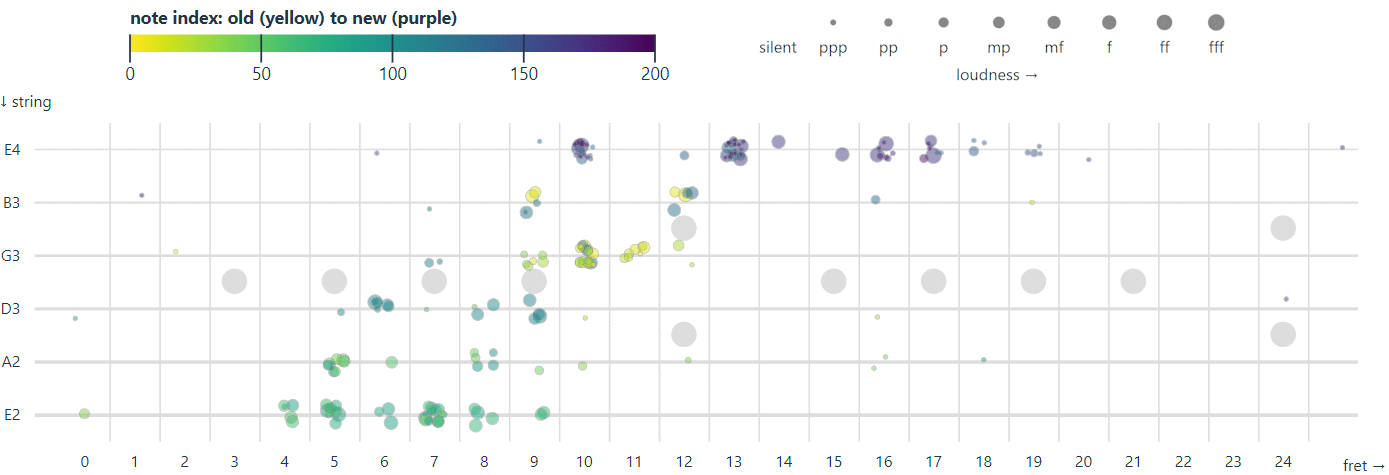}
      \caption{%
      	\chgmin{}{A scatterplot of played notes, colored by temporal order.}%
      }
      \label{fig:fretboardjitter}
    \end{figure}

\paragraph{Design}
Our design process started \chgmin{from the fretboard as shown above and simplifying it into a Cartesian}{by simplifying the fretboard into a} grid\footnote{\chg{}{Fretboard: \href{https://visvar.github.io/mila/?d=fretboard-jitter}{visvar.github.io/mila/?d=fretboard-jitter}}}.
We draw each played note as a dot in a scatterplot at the (fret, string) position it was played.
To diminish extra notes (noise), we encode loudness as the dot's area, and to reduce overplotting, we apply jitter and semi-transparency.
We encode the temporal order of notes through a color gradient.
\chgmin{Below}{In \Cref{fig:fretboardjitter}}, we can see a diagonal movement from bottom left to top right.

While this scatterplot and color encoding work for short segments, they do not scale for full improvisations.
We address this limitation by only showing the last few bars (user-adjusted) and through a small multiple approach, where each facet shows one bar (or more).
An alternative coloring dual encodes the fret with a multi-hue color scale, improving pre-attentive pattern detection but losing temporal information within facets \chgmin{(see below)}{(\cref{fig:fretboardfacets})}.

\paragraph{Examples}
\chgmin{The image below}{\Cref{fig:fretboardfacets}} shows an improvisation that starts with horizontal playing and then jumps to another position in the second bar.
Because the colors are identical, we see that this position contains the same notes (one octave higher), which would require close reading without coloring.
In bars 3 and 4, the guitarist switched to horizontal playing and used the open strings (red dots at fret 0).
The last bar was again played with the same position as bars 1 and 2, indicating that the musician should learn different ones.

    \begin{figure}[htb]%
      \centering %
      \includegraphics[width=0.7\linewidth, alt={Five scatterplots with the same layout as above. The first and second show clusters at different positions, but the dots' colors are similar. The third and forth plot's dots are streched out horizontally. The fifth resembles the second.}]{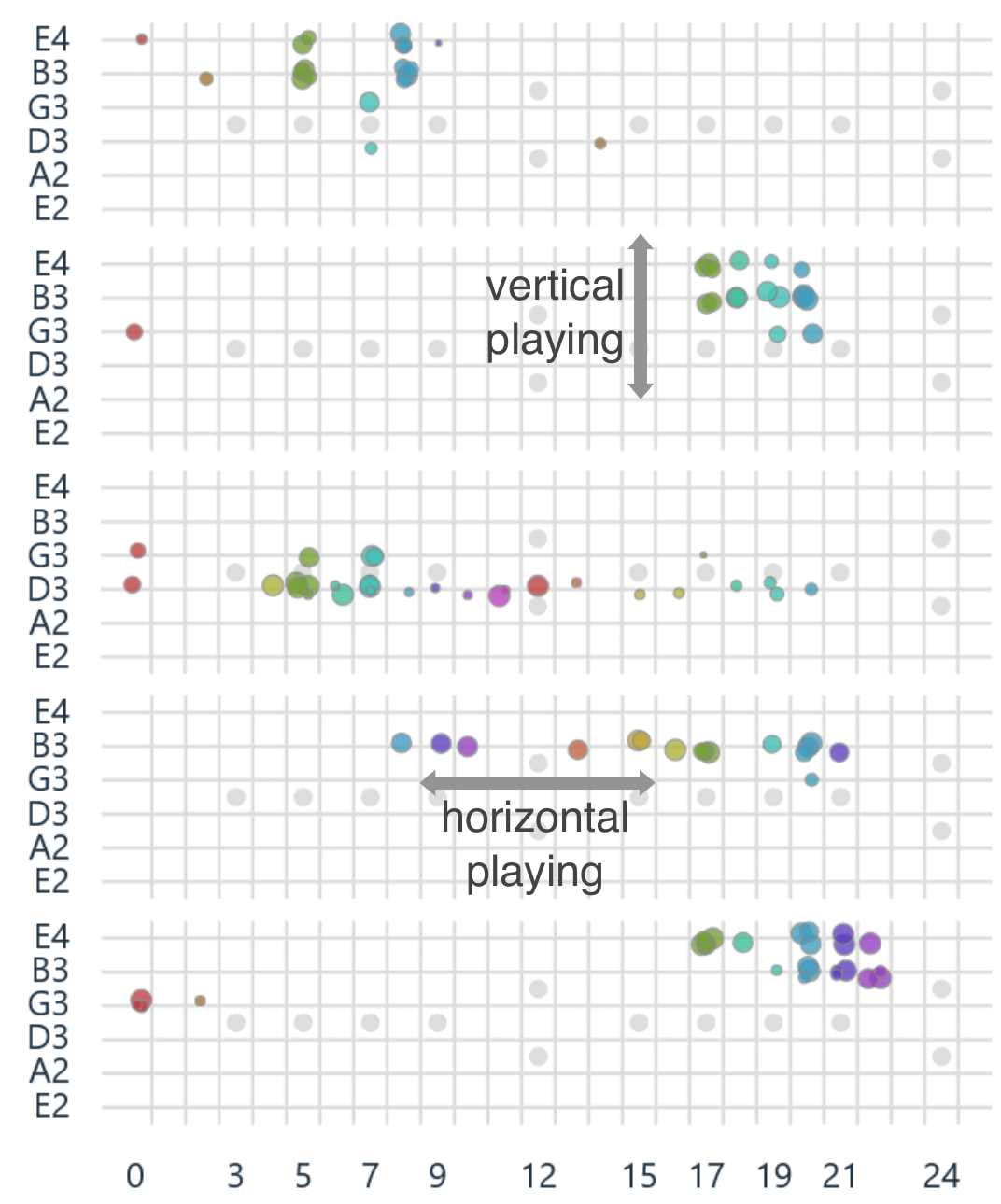}
      \caption{%
      	\chgmin{}{Facets per bar show different playing directions.}%
      }
      \label{fig:fretboardfacets}
    \end{figure}

\paragraph{Take-Away}
1) Encoding time through color and facets works for short and longer time spans respectively.
With few bars per facet the visualization takes space and requires looking around, with more bars each facet gets cluttered and might show a mix of multiple patterns -- the optimal setting depends on how long the improvisation is and how much it moves around the fretboard.
2) A simple dual encoding of position with color can reveal patterns that would be hard to see otherwise, as shown by the above example, where the same position was shifted by 12 frets (one octave).

\section{\chgmin{Evaluation with Musicians}{Preliminary Feedback on Designs}}\label{sec:expert_evaluation}

To further evaluate our designs, we conducted a qualitative study with musicians who have not been part of the development process.
The overall goal of this evaluation was to confirm whether our visualizations indeed show unhearable patterns that are relevant for learning -- that allow musicians to get the insights they seek.
A second goal was to learn \emph{how} such designs might be used in music education.
\chgmin{}{%
We adopted a qualitative methodology, as our aim was not to assess the effectiveness of the specific designs, but to understand the effectiveness of the approach more broadly.
}
For participants, we sought music learners who are the potential users of our tools, but also music educators who could provide more perspective on how tools may fit into the learning process.
The data we collected consists of answers about demographics and prior experience, observation of their visualization usage, quotes, anecdotes, and answers from a semi-structured interview.
See our supplemental material for more details on our evaluation.

\subsection{Research Questions}

    \emph{Can musicians see unhearable and relevant patterns?}
    The~first question we wanted to answer is whether regular musicians can see patterns in the visualizations that 1)~would be hard to hear and 2)~they find relevant for their practice -- not only showing patterns but actually improving or supporting the learning process.
    During the study, we are looking for signs that, after explanation and examples, the participant can analyze their own playing.
    
    \emph{How would our designs be used in learning and teaching?}
    The second question we investigated is how our visualizations would be used during practice and how teachers would integrate them into their lessons or homework. 
    Specifically, we want to learn about the context in which they would be used (situation, people, time of usage, time of analysis), as well as about constraints and limitations.

\subsection{Participants}

    We recruited 13 participants through convenience sampling (P1--P13). 
    One identified as female, the rest as male. 
    Five were 18--29 years old, five 30--39, two 40--49, and one 70--79. 
    All had normal or corrected vision, except P3 who has a color-vision deficiency.
    They had varying levels of experience with different instruments and played between 3--60 years (mean: 21.92).
    For self-assessment, our participants record \chgmin{themselves}{their practice playing} never (1 participant), rarely (6), sometimes (2), regularly (3), or often (1).
    They use visuals never (10), rarely (1), sometimes (1), or often (1).
    Some mentioned using tools like digital audio workstations (DAWs), Synthesia (a~piano learning app), and tuners to visually assess their playing.
    Our participants had received \chgmin{instruction on instruments}{instrument lessons} for 0.5—25 years (mean: 9.27).
    Seven had teaching experience, three of them taught for 5--29 years to paying students (P3, P4, P12)
    and three have degrees in music or music education (P4, P5, P12).

\subsection{Methods}
    We conducted the study both \chgmin{online via meeting software}{through video calls and screen sharing} (5 participants) and in our lab (8). Online studies allowed us to recruit participants from a wide range of places, while lab studies allowed us to provide MIDI instruments.
    After giving consent, the participant filled a short demographic survey asking about their experience with instruments, with 
    learning 
    and teaching music, their familiarity with visualization, and experience with recording and visual feedback. 
    We recorded the screen together with the desktop and microphone audio, and we prompted them to think aloud. 
  
    For the practical part, we first asked them for current goals and drills they work on, and what they have trouble with hearing.
    We did not expect perfect matches between the drills we support and the participants' goals, but tried to identify designs that were close enough that they could appreciate what they may provide for someone working on a drill it was designed for.  
    Each session started with a simpler design and encouraged the participant to first try basic playing and then an exercise or excerpt of a piece they would usually practice or teach.
    We also instructed them to try out different playing, for example, to experiment with different rhythms. 
    Other than that, participants were free to play whatever they liked or were interested in getting feedback for.
    Looking at the visualization together, we waited for them to interpret it and only helped where necessary.
    The MIDI-capable instruments they used during the study were keyboard (6), guitar (3), drum (2), and saxophone (1).
    Three participants tried our audio-based designs with singing. 
    P4 did not have access to a MIDI instrument, so we showed them examples from others.

    After the practical part, we conducted a semi-structured interview.
    Participants with teaching experience were asked additional questions on how they would integrate them into lessons and how they would guide/advise students to use them.
   
    Each participant's session took about 2 hours on average (ca. 25 hours total).
    The study was approved by our university's ethics commission (application No. 22-039) and participants were compensated 30 \chgmin{Euro}{euros}, except six who, as employees of the university, were not allowed compensation.

\subsection{Results}

    In the following, we summarize the results of a thematic analysis of the video/audio recordings of the participants' visualization usage and interviews.
    We structure the results by themes ordered along our research questions.

    \subsubsection{Can musicians see unhearable and relevant patterns?}

            \textit{Seeing patterns.}
            All participants saw at least some patterns they did not hear while playing.
            We identified different aspects that are hard to hear.
            For example, hitting a specific loudness or tempo can be hard to confirm by hearing (P12, P13).
            Small deviations are hard to hear, especially for beginners: ``pupils have difficulty listening to themselves and perceive differences'' (P5).
            Still, even small deviations can be relevant: ``if I was a musician on stage, every note would matter'' (P13).
            Another difficulty are mixed patterns,
            when multiple notes or drums are played together (``with the hi-hat at the same time'', P6).
            P7 also struggled to integrate over time: ``especially having these several bars [where] I missed the metronome more in the beginning then I do now, that’s something I wouldn’t audibly remember ... the history of your playing is also something I wouldn’t hear''.

            \textit{Relating feedback to what is played.}\label{evalresult:relatingfeedbacktoplaying}
            Relating patterns in a visualization to what has been played can be difficult at times.
            For instance, while using the chord progression design, P12 assessed that ``for the drill it is good, but it is hard to infer anything from my playing''.
            Furthermore, P5  found ``it hard to retrace when you play so much at once'', showing a limitation of some designs to become less legible with an increasing amount or complexity of data.
            Real-time feedback can help here: continuously watching the visualization makes relating actions to feedback much more direct than recording and listening ``you don’t know how exactly how you sang before. Such a direct visualization is definitely better'' (P8).

            \textit{Relevance for practice.}
            Our participants found at least some of the patterns they saw relevant for their own or their students' practice and that the visualizations ``help to get control'' (P1).
            They also found use cases we did not intend (P1--P5, P7, P8, P11), such as songwriting, learning about the intervals in chords, and composing music that also \emph{looks} \chgmin{good}{beautiful} in a visualization.

    \subsubsection{How would our designs be used in learning/teaching?}

            \textit{Need for visual support.}
            We expected teachers to be critical and negative about visual tools that could detract from learning to listen.
            P5 agreed that some teachers might be negative toward such tools, also because they ``don’t want to make themselves dispensable'', as giving feedback is part of what they are trained for. 
            Our participants mentioned the need for feedback during practicing without a teacher, ``do I move blindly into a wrong direction?'' (P8), found that ``through the visualization, [self-assessment] is approachable, comprehensible'' (P12), and concluded that the designs are ``really useful as a learning tool for both me and my students'' (P3).
            P4 told us that our designs would \chgmin{be useful for}{help} professional musicians who have to put in more time ``this would be helpful for people like me who are serious'' and for students who are examined by a school or university: ``for everybody going through exams [...] that would be super helpful''.
            P4 thought that ``in a school system or a graded system, these are excellent because you would see if \chgmin{they}{[students]} are doing the exercise correctly''.
            
            \textit{Combination of hearing and visualization.}
            However, P5 believed that a combination with visualizations can be better than only listening, because ``the interplay could be valuable. When you have both, the visual and the auditory [...] some can work better with the visual'' and P4 stated that ``a lot of these apps would be a great addition to any kind of ear training or theory or musical skills class''. 
            Still, in self-assessment, ``a large part should be listening to oneself and [assessing] how it sounds and if it’s good'' (P12).
          
            \textit{Need for guidance.}
            \chgmin{Designs would first need to be introduced one by one (P1).}{}
            P12 suspected it to be difficult for students to find out which design to use and how, ``but if I’m working with someone on something specific [...] I could show them during the lesson how to work with it and what the criteria are they should pay attention to [...]''.
            We note that our current prototypes serve exploration and would be simplified.

           \textit{Time and scheduling.}
            Practicing with visual feedback could take up ``10--15 minutes during the lesson'' 2--3 times per week (P1), or even ``50\% of what they do or 75\%'' for school students (P4).
            P5 would use the visualizations at the start or middle of the lesson, and at the end rather \chgmin{than}{} let them train their ear, whereas P12 would not use it much during class but rather during homework. 
            
            \textit{More efficient assessment.}
            Having a better idea of how students did between lessons could make teaching more efficient: ``If a student comes into a lesson and they’ve sent me 3 screenshots of an accent exercise and they nailed it, I’m not going to waste time having them play it in the lesson ... we just go [...] to the next one'', P4 said; this would ``save a lot of time in a lesson''.

            \textit{Looking on-the-fly or post-hoc.}
            Some participants mostly avoided looking at the visualizations while playing (P1, P13), others did that at least for some. 
            Whether watching is practical depends on how difficult the playing is -- playing the same note and only changing rhythm or loudness does not require as much focus or looking at the instrument as improvising.
            With some designs, P7 ``could really play and look at the same time and understand how I should change in real time [...] while in some others, when I was playing, it disrupted me.''
            They concluded ``it was all about learning when to look and when not''. \label{evalresult:learnwhentolook} 
            P11 found that ``looking at the screen distracts me a bit when I don’t know exactly what to play [...] it depends on what I play and how well I practiced that''.
            P4 ``would recommend that [students] start by watching their screen as they’re playing, and once they are getting close to where they want to be [...] then you try without looking, and then you can compare the two visualizations. And see ... are you getting better?''. 
            They liken this approach to using ``training wheels and then you take them off''.
            P12 stated that it ``makes more sense to focus on oneself during playing [...] to goal is later to be able to play without app [...] first get your own impression [...] then get feedback and then, based on that, question your opinion and maybe strengthen your own perception ... and in further iterations try to integrate both the app’s feedback and your own hearing experience''.

            \textit{Usage of parameters.}
            Some participants \chgmin{(P1, P2, P3, P4)}{} were eager to try out different visualization parameters, often without us telling them to. 
            They, for example, changed the amount of data/time shown (P2) or coloring (P3), or used playback (P4, P12). 

            \textit{Comparison.}
            P4 and P5 would show students their own visualized playing to show them how it should look and that it does not have to be perfect.

         \textit{Isolation/combination of elements.}
            P1 told us that students might have problems with things that feel easy to more experienced musicians, and that one challenge is to simplify things to help them learn and \chgmin{not lose motivation}{stay motivated}.
            While isolating elements can help focus on them, P5 stated that for some cases, it \chgmin{would be important}{helps} to see multiple elements together, such as off-beat and duration.
            In our Accents design, we show loudness and timing combined, which already allows to find correlations: ``I don’t know if a made a trade-off between precision and how loud I play accents but it looks like it'' (P12).

        \subsubsection{Further findings}

            Besides answers to our research questions, we got further insights from our participants, such as potentially interesting design dimensions we did not explore.
            Our participants felt that our collection of designs is already mostly exhaustive: ``already contains already a lot of what you can consider and practice [...] this is already very all-encompassing'' (P12).
            P7 especially appreciated the flexibility to play almost anything, not only along a song or exercise, and get feedback for it.
            Among their suggestions for additional designs were more aggregated results for improvisation (P2) and an encoding based on the circle of fifths for a chord progression \chgmin{``seeing, does it fit in there''}{to help see} ``does it fit in there'' (P7). \label{evalresult:moreaggrationwanted}
            Others had a few ideas for what to add beyond our intended scope, such as gamification, scores for tracking progress, and comparison to sheet music. 
            P11 told us that \emph{seeing} how you play already motivates you, similar to gamification (``I can do this better!'').
            P2 was interested in the timbre of their singing (the acoustic spectrum) to see ‘spaces’ in which one sings, such as chest voice and head voice, which they find ``super hard to hear''.

        \subsubsection{Takeaways} 
        We were encouraged by the study results. 
        They suggest that visualizations are able to ``make the unhearable visible'' and, more importantly, this can be useful in musical practice. 
        We are encouraged by the emergent use cases: participants found value in the tools beyond what we had originally designed them for. 
        Educators were able to envision many roles for such tools. 
        However, the \chgmin{study }{}results also suggest challenges that must be addressed in specific designs and tools. 
        Broad challenges, such as musicians dividing their attention between playing and watching or managing the complexity of the data in longer exercises, must be considered in created designs that will work in practice. 
        While the experience with our initial prototypes is encouraging, it does suggest a number of design considerations as we develop actual applications.

\section{Design Considerations}\label{sec:considerations}

In this section, we discuss considerations for the design of visualizations that support musical instrument practice. 
\chgmin{}{%
These considerations are based on what we learned during the whole design process.
}
Instead of requirements or recommendations, they are meant to guide thinking about options and trade-offs for specific circumstances and are by no means complete.

\subsection{Can Automatic Assessment Replace Visualization?}

The first consideration for any visualization is whether it is even needed or if there is a more automatic solution.
While there has been work on AI for textual practice feedback~\cite{shu2025fretmate}, we argue that it should augment but not replace visualization -- assessment, such as scores or text, is just the most abstract overview, a good start for knowing when or what to analyze. 
Visualization then provides other levels of abstraction with more context down to the full details.

Assessment of musical practice is hard because there are no general criteria for what is good.
As an example: what is the threshold for good enough timing in rhythm? 
It could be $x$ seconds, $x$ percent of a note duration, or even something non-linear. 
Would it depend on the tempo, instrument (or part of it for drums), or the musician's skills and goals?

\begin{figure}[htb]
  \centering 
  \includegraphics[width=\columnwidth, alt={A left-right arrow that points to `system knows more' and `system knows less' with examples of known aspects written above the arrow.}]{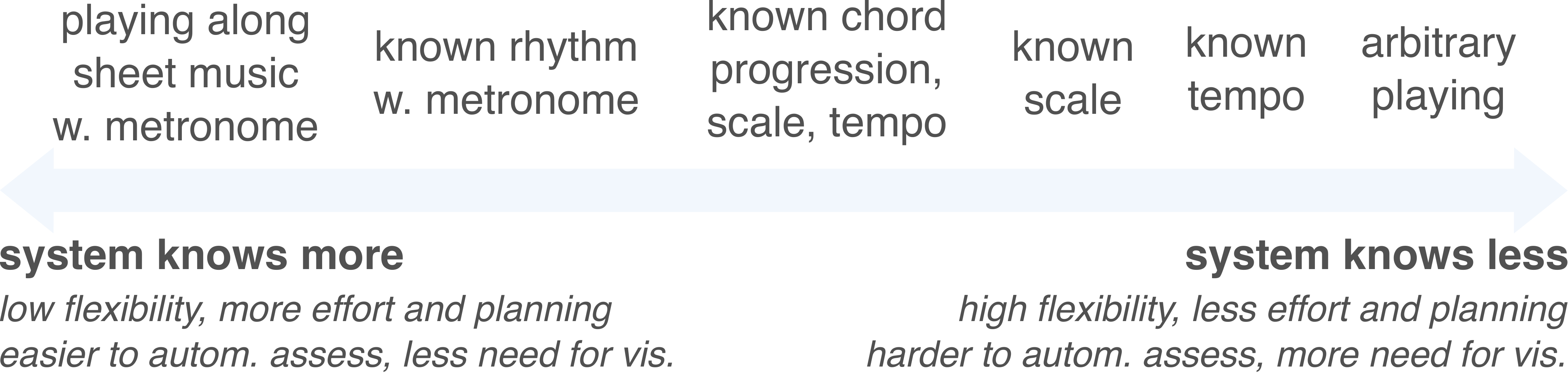}
  \caption{%
The more a system knows, the easier it is to give feedback: when we know the exact notes the musician should play, we can give feedback on every single one.
Conversely, that means that the more flexible a design is, the harder it is to automatically assess playing.%
  }
  \label{fig:flexibility}
\end{figure}

This dependency also brings a trade-off between flexibility and utility.
There is a complex spectrum of how much the system knows (\cref{fig:flexibility}).
For example, in our improvisation design (\cref{sec:design-chordprog}), \chgmin{the system}{it} needs to know the chord progression.
Knowing what the user is trying to do is hard: a practicing musician might start and stop arbitrarily, repeat pieces over and over, or play a simplified version of some material.
A good teacher can listen to bits and pieces of a song and figure out the appropriate context.
With a system, we have to decide how much we want to force structure on the practice.

Scores like ``good'' or ``80\% correct'' are not transparent when the definition of good and correct is not understood by the user.
Metrics can be easy to understand when they are visualized: our notes-within-tolerance percentage (\cref{sec:design-timing-consistency}) has a visual representation of ticks that fall inside areas.
While a metric can detect that the playing is off-beat, and indicate how much in either words (slightly) or numbers (0.1 beats), a visualization can also indicate where this is happening by showing it in the context of a baseline such as a grid or even someone else's data.

\chg{}{%
    More generally, visual comparison can help assess what is good or how one could improve when there is no well-defined metric available.
    For example, comparing multiple repetitions cannot show whether each one is good, but reveals how consistent and, therefore, controlled the playing is.
    Comparing to the recording of a teacher can show what range of deviation from a rhythm is acceptable.
    For improvisation, seeing how the playing of others looks can help inspire, by using different intervals, using them in a different order, or using the instrument differently.     
}

Automatic assessment might make mistakes (hallucinate), have different assumptions than the musician on what matters (misalignment), and be less robust against systematic and random noise.
In these cases, visualization helps to double-check, can (interactively) show multiple levels of strictness, and can make uncertainty in assessment explicit.

In short, consider for each case whether 
1) automatic assessment is sufficient or too limited by ill-defined and diverse requirements and 
2) whether it could be better \chgmin{communicated to the musician}{conveyed} through a visual second channel and more context.

\subsection{Which Unhearable Things are Worth Making Visible?}

After deciding \textit{whether} to visualize, one needs to consider \textit{what} to show.
Showing only relevant patterns is a common challenge in visualization.
Unlike many other applications, our data can already be perceived (by hearing), refining this challenge:
Patterns that cannot be heard even by an audience usually do not matter, except, for instance, for dedicated technique exercises (\cref{sec:design-timing-consistency}).
Conversely, patterns that are obvious through hearing are less important to visualize but can serve as context.

Therefore, the main challenge in giving feedback is the robust determination of not only what is good or bad, but also \textit{relevant}.
Finding a balance between showing unnecessary details and filtering away the signal is difficult, especially for a diverse and ill-defined application such as music practice: what is relevant depends on the instrument, personal goals, and more.
If we err on the side of showing too little, the musician can focus on the most important issues first but might miss others.
When showing too much, we overburden the user, and they might practice the wrong things such as minimizing deviations that are already good enough.
However, some visible issues that strictly cannot be heard might still be relevant for proper technique and, for example, serve as a hint for better hand position/movement (\cref{sec:design-timing-consistency}).

Designers can consider different strategies to combine overviews and details or find a good compromise:
1) reducing details by rounding, binning, or discrete encodings like note symbols that hide small mistakes (\cref{sec:design-accents}),
2) showing multiple levels of detail at once, such as aggregation and raw data or bar chart and quantized symbols,
3) highlighting issues visually -- or, to help motivation, making ``good enough'' explicit, for example, with tolerance areas (\cref{sec:design-timing-consistency}).

\subsection{How Can We Use Existing Mental Models?}

For every visualization, one should consider existing mental models to facilitate onboarding, understanding, and \chgmin{transfer}{transferring} of insights via familiar concepts~\cite{liu2010mentalmodels}.
One example is using musical units in visualizations: musicians think in beats or quarters instead of seconds, in note chroma and cents rather than Hertz, and in pianissimo or forte, not decibels.

More interestingly, music notation and education already bring many visual elements we can directly transform into layouts and encodings. 
For instance, common sheet music notation lays out time from left to right in bars over multiple rows.
Within each bar time is not encoded linearly but compressed and quantized through note symbols, which encode durations.
Pitch is encoded through these symbols' vertical position and additional modifiers like \musSharp\ and \musFlat.
Besides this notation, other forms of sheet music exist, such as guitar tablature that resembles the instrument with one line for each string on which notes are displayed by their fret number instead of pitch.
Another visual element in music is the physical layout of an instrument, where notes have spatial locations.
Educational material contains visual representations such as the circle of fifths and the duration pie charts we based one design on (\cref{sec:duration-pies}).

Existing representations, however, are not intended to show imperfect and usually messy practice data.
Here are considerations on deconstructing, adapting, or abstracting them:
One extreme would be using sheet music exactly as is with visuals drawn on top~\cite{park2025musicdynamics} or a guitar fretboard with frets spaced like on a real guitar.
These might be close to what musicians know, but not optimal for perception: writing dynamics such as $p$, $m$, or $f$ on top of notes is harder to perceive than symbol size.
A slightly abstracted fretboard with equal fret spacing is easier to read, more space-efficient, and already used in chord diagrams and guitar education.
Since note symbols are not intended for fine timing, they can only encode certain values (that are not spaced \chgmin{equally}{uniformly}) without becoming visually complex (\cref{sec:design-accents}).

Designers have to consider the goal and task -- when comparison is important, abstract bars might be better than familiar symbols because position is easier to perceive than size.
Visual scalability is another factor~\cite{richter2024scalability}. 
Symbols need more space and are harder to perceive than abstract encodings, especially when there are many notes shown.
Another consideration is the individual musician.
For example, many guitarists use tablature instead of common sheet music notation and might not be familiar with reading note symbols.
On the other hand, a visualization that uses them could incentivize them to learn a potentially useful concept.
Similarly, visualization can teach alternative mental models and help thinking visually, for instance in the case of positions one uses (or has not used yet) while improvising on a guitar.

\subsection{How to Trade Off On-The-Fly Versus Post-Hoc?}

We explored both on-the-fly use of visualizations and post-hoc analysis.
Immediately showing what is played lowers the bar to entry:
Live visualization provides immediate benefits and to directly adjust the playing or visualization and to spontaneously try out different exercises without having to plan or pause.
\chgmin{}{Our participants also found it easier to relate specific actions to visuals when seeing them live (\cref{evalresult:relatingfeedbacktoplaying}).}
On the other hand, live feedback will be overwhelming in many situations as it distracts from focusing on the instrument.
However, paying constant attention is not required, as a musician can take a glance whenever interested.
For reflective analysis, specifically, right after finishing an exercise, we use a display that stops moving once the musician pauses.
The combination of different encodings or an adaptable level of detail allows glancing~\cite{blascheck2021characterizing} or incidental usage~\cite{moreira2024incidentalvisualization} while playing and looking at more detail whenever there is interest.
In our evaluation, we found that musicians have to learn when to look and when not \chgmin{}{(\cref{evalresult:learnwhentolook})}, and that the practicality of watching while playing depends on what is played and on which instrument.

\begin{figure}[tb]
  \centering 
  \includegraphics[width=\columnwidth, alt={A plot with thumbnails of our designs postitioned by typical time scale (X axis) and observation mode (Y). Below the X axis typical musical time scales are shown as rectangles.}]{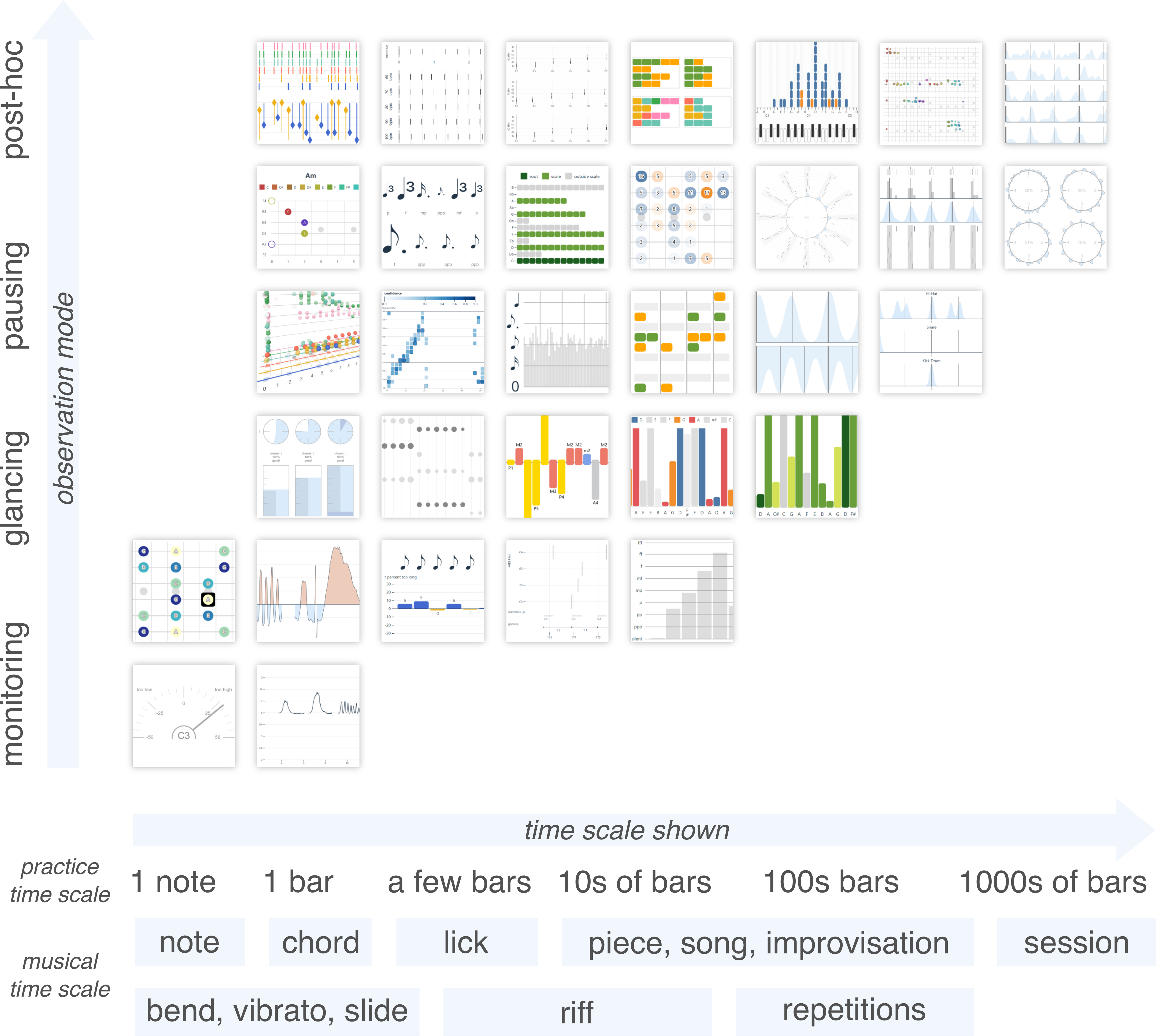}
  \caption{%
  	The typical time scale and observation modes for our designs.
    Positions are approximate and designs cover ranges, not single points.
    We can see a correlation: monitoring is most appropriate for small time frames such as one to a few notes, chords, or bends, whereas more complex scenarios on longer time scales, such as improvisation, are more suited for glancing, pausing, or post-hoc analysis.%
  }
  \label{fig:time-realtimeposthoc}
\end{figure}

There is \chg{}{also} a relationship between how much data a visualization shows and the observation mode (\cref{fig:time-realtimeposthoc}).
Live monitoring is most useful for issues that can be quickly reacted to, for example, pitch bends that can be corrected even within the current note.
For patterns that stretch over longer time scales, such as tempo keeping, more data needs to be shown and sporadic glancing suffices.
Comparing multiple repetitions or looking for variance in improvisation requires pausing\chgmin{ for a moment}{}.

\begin{figure}[tb]
  \centering 
  \includegraphics[width=\columnwidth, alt={A similar plot to the one above. This time the axes are time aggregation (X axis) and value aggregation (Y axis).}]{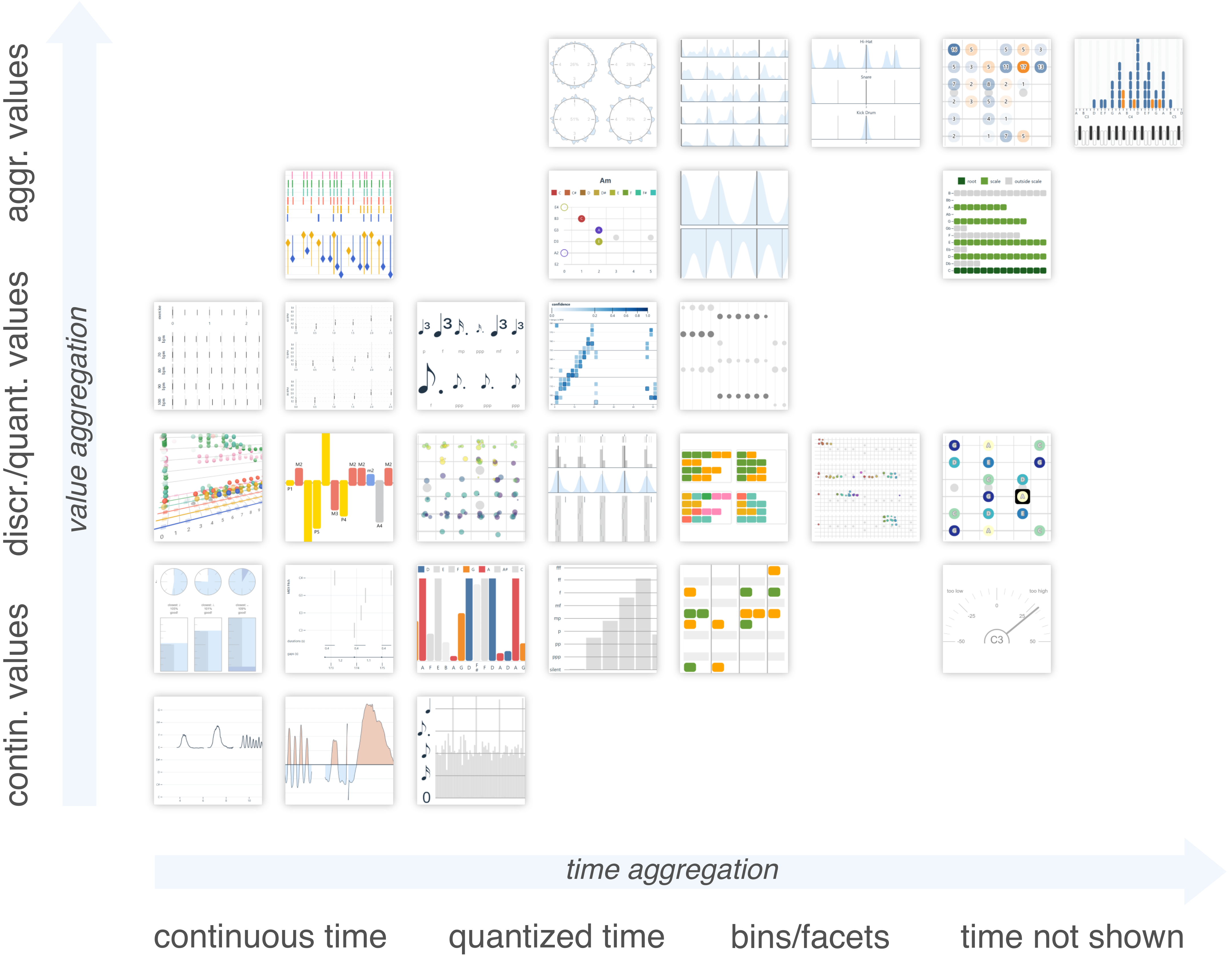}
  \caption{%
    Our designs cover many different combinations of time and value aggregations, but leave gaps for \textit{continuous time \& aggregated values} and \textit{time not shown \& continuous values}.%
  }
  \label{fig:aggregation}
\end{figure}

\chgmin{%
    [moved to \cref{consid-scalability}]
    For larger time scales, aggregation is more important as it allows abstracting data to a level that makes it easier to summarize and compare.
    Time and other values, including those derived from time such as note durations or tempo, can be aggregated separately in different ways (\cref{fig:aggregation}).
    For example, we facet time by bar and repetition and quantize durations in our chord progression design (\cref{sec:design-chordprog}). 
}{
}

\chgmin{%
    As music practice visualization is likely used in personal~\cite{huang2015personal:vis:and:va} and casual~\cite{pousman2007casualvis} settings, the effort has to be worth it -- a good user experience is critical unlike in professional cases. 
    An easy setup, minimal interaction, and on-the-fly usage save time and effort before and during analysis.
}{}

\chgmin{}{%

    \subsection{How Can Visualizations Scale With Larger Data?}\label{consid-scalability}

    As the number of notes an instrument will produce in a short time is limited, visualizations can be easily designed to accommodate the expected amount.
    Therefore, scalability challenges mostly concern the duration of recorded playing.
    %
    During instrument practice, a learner might want to look at playing of varying duration depending on how much context they need.
    For example, when repeating an exercise many times, looking at all repetitions together helps analyze the consistency or progress.
    Longer recordings usually contain more notes and can therefore pose a challenge to the visual scalability\cite{richter2024scalability}.    
    Musical instrument practice brings a special scalability concern, because the learning modality and analysis task likely changes when playing longer passages:
    When repeating a short pattern, one can look into individual note timings.
    At the scale of a full song or improvisation, it rather is the overall shape of tempo consistency/variation that matters.
    
    In our design exploration, we found that for larger time scales, aggregation is more important as it allows abstracting data to a level that makes it easier to summarize and compare.
    Time and other values, including those derived from time such as note durations or tempo, can be aggregated separately in different ways (\cref{fig:aggregation}).
    For example, we facet time by bar and repetition and quantize durations in our chord progression design (\cref{sec:design-chordprog}), but one participant wanted it even more aggregated (\cref{evalresult:moreaggrationwanted}).
    
    There are different kinds of scalability in music practice:
    the number of notes per time unit,
    the length of the practiced musical pattern,
    the number of repetitions or takes,
    and the number of instruments and musicians.
    We have not addressed the last one and there are likely more that we did not yet consider, which are opportunities for future work.
}

\chgmin{%

    \subsection{How Can We Evaluate Design for Practice Support?}
    
    Ideally, we would like to directly measure the quantitative impact of tools on learning.
    This is hard because it requires quantifying learning effects such as an increase in skill that are highly individual.
    Learning effects go beyond performance metrics like ``number of mistakes''; the more important question is if a tool helps make better music, and critical for a personal use case, if using it is enjoyable.
    Furthermore, controlling for confounding factors is hard.
    For example, it might be unclear which part of a system leads to improvement; it could be a novelty effect that led to more effort spent on practice but wears off.
    
    In this work, we focused on exploring what is possible instead of evaluating a specific solution.
    Similar to a standard design study~\cite{sedlmair2012design}, we already contain some evaluation in the process through formative expert feedback and design critiques by presenting ideas and prototypes within our institute and demonstrating them to guests.
    We characterized the designs in tables regarding data, skills, and design patterns to identify gaps worth exploring further.

    Since it is hard to evaluate this collection of designs with only one way, we ``triangulated'' the evaluation from different angles:
    Through auto-biographical design, we made sure that our collection is useful at least for a small group of musicians. 
    Additional feedback from participatory design allowed us to extend and improve this collection.
    We then evaluated whether musicians who were not part of the design can see relevant patterns in the visualizations.
    Last, we also interviewed some of these musicians, who have teaching experience, about using them in their teaching.
    Our case studies and evaluation with musicians demonstrate the broad utility/validity of visualization, but we did not directly evaluate usability and adoption.

}{}

\subsection{Further Considerations}

We did not explore feedback modalities other than visualizations shown on a screen.
For some exercises, information shown directly on the instrument via augmented reality and situated/embedded visualization~\cite{willet2017embeddeddatarepresentations} would help to directly relate feedback to where the notes came from.
For example, our fretboard-based designs (and adaptations to other instruments) could be directly displayed on or near the instrument.
Depending on the display hardware (head-mounted, projected), encodings can make use of all three spatial dimensions, for example, to encode 2D space and time in a space-time cube \cite{heyen2022cellovis}.
Embedded visualization could also be used as an overlay to video, either for the above instrument-based visualizations or to provide feedback on the musician's hand or body pose, which are not only relevant for playing but also to avoid strain and injury.     
In our study, P9 mentioned auditory feedback or haptics, which would allow musicians to not look at the screen but focus more on their instrument.
Both these modalities could interrupt or distract the user, as they are timed by the system and are harder to ``ignore'' than visualizations, where users can pause and glance anytime they want.
Auditory feedback would further have to be designed in a way that does not interfere with listening.
A sub-field of sonification research uses music-like encodings~\cite{roennberg2019musicalsonification}, which \chgmin{pose}{are} an interesting opportunity to explore for future work in the context of instrument practice.
Most promising would be a combination of such modalities.

\chgmin{%
We focused our exploration on the more abstract MIDI data instead of audio, which brings benefits in terms of data acquisition and processing, but also limitations.
Of course, only audio contains the full richness of what can be heard.
While there are ways to get continuous data from MIDI (we noticed in our expert study that the MIDI saxophone of P9 outputs loudness continuously instead of once per note), not all instruments support this.
Furthermore, MIDI cannot provide information about the timbre, which is important for singing, as it represents how a note sounds.
}{}

\section{Limitations and Future Work}

\chgmin{%
    Our exploration is limited to MIDI-capable keyboards, drums, and guitars (except for designs using audio pitch).
    The more general designs can be easily extended to any instrument for which reliable real-time MIDI conversion exists.
    For example, MIDI saxophones and xylophones are available and there is work on software \href{https://www.jamorigin.com/products/}{for violin}.
    Without discrete notes, abstraction is more difficult and visualization is often limited to line charts~\cite{wang2021soloist} and waveforms~\cite{park2025musicdynamics}.
}{%
    We focused our exploration on the more abstract MIDI data instead of audio, which brings benefits in terms of data acquisition and processing, but also limitations.
    Our exploration is limited to MIDI-capable keyboards, drums, and guitars (except for designs using audio pitch).
    The more general designs can be easily extended to any instrument for which reliable real-time MIDI conversion exists.
    For example, MIDI saxophones and xylophones are available and there is work on software for violin\footnote{\href{https://www.jamorigin.com/products/midi-violin/}{jamorigin.com/products/midi-violin/}, accessed Jan 5, 2026}.
    Second, only audio contains the full richness of what can be heard, including the timbre that represents how a note sounds and is especially important for singing.
    Still, for instruments like keyboards and e-drums, MIDI is sufficient to reproduce the sound almost perfectly.
    Future work could explore more abstract visualizations also for audio data.
    However, abstraction is more difficult without discrete notes and therefore audio visualization in related work is often limited to line charts~\cite{wang2021soloist} and waveforms~\cite{park2025musicdynamics}.
}

We focused on individual practice.
Future work could include comparison to a teacher’s recording or data of others practicing the same exercises, which makes (automatic) assessment easier.
Taking a step further, one could even compare band members who are playing simultaneously in separate voices with different instruments.
For more professional use cases, more complex interfaces and long-term data collection would extend the space of appropriate designs.
\chgmin{}{Such new directions will bring new challenges, for example for scalability.}

Our work is a first step in music practice visualization that shows possibilities and potential.
\chgmin{}{%
    Therefore, our evaluation aimed to confirm that visualizing unhearable patterns helps in musical instrument practice, and the high-level feedback we received is encouraging.
}%
Future evaluation is required to learn which designs musicians accept.
We plan a \chgmin{}{long-term} field study where we record how people use our online prototypes.
Such a study is limited, due to a bias toward tech-savvy musicians and because some features need specific soft-/hardware, such as a MIDI pickup for fretboard visualizations.
\chgmin{}{%
    In our evaluation, most participants where intermediate or expert musicians. 
    Future studies should include a broader audience of more ordinary users and a smaller set of visualizations that require less musical expertise.
    Such studies would allow for a larger number of participants and more robust and objective methods.
    A quantitative and task-based evaluation of individual design components would tell what encodings support different data and tasks best.
}

\section{Conclusion}

We created a collection of designs to explore how visualization can support musical instrument practice by making unhearable patterns visible.
We believe that the designs and trade-offs we describe and the considerations we discuss inspire and support future work and products.
The prototypes we created are publicly available as a web app and source code and we plan to extend them. 
\chg{}{An initial study suggests that the approach of visualizing practice data provides value to learners and has a variety of uses in music education.}
A future field study and survey among (potential) users will help evaluate adoption and real-life utility.
We further hope to inspire researchers in domains such as machine learning and music information retrieval, as well as hardware manufacturers, to consider real-time feedback for musicians as a new use case. 
New and better techniques will enable more diverse and effective visualization for instrument practice.

\section*{Acknowledgments}
This work was funded by the Cyber Valley Research Fund 
and supported in part by NSF (USA) award 2007436.

\bibliographystyle{abbrv-doi-hyperref}

\bibliography{bib}

@String{BIT = "{BIT}" }

@String{Computing = "Computing" }

@String{Computer = "{IEEE} Computer" }

@String{Academic = "Academic Press" }

@String{Springer = "Springer-Verlag" }

@ARTICLE{chang2023muse,
  title={{MUSE}: Visual Analysis of Musical Semantic Sequence}, 
  author={Chang, Baofeng and Sun, Guodao and Li, Tong and Huang, Houchao and Liang, Ronghua},
  journal={IEEE Trans. Visualization and Computer Graphics (TVCG)}, 
  year={2023},
  volume={29},
  number={9},
  pages={4015-4030},
  doi={10.1109/TVCG.2022.3175364}
}

@ARTICLE{willet2017embeddeddatarepresentations,
  title={Embedded Data Representations}, 
  author={Willett, Wesley and Jansen, Yvonne and Dragicevic, Pierre},
  journal={IEEE Trans. Visualization and Computer Graphics (TVCG)}, 
  year={2017},
  volume={23},
  number={1},
  pages={461-470},
  doi={10.1109/TVCG.2016.2598608}}

@article{roennberg2019musicalsonification,
  author    = {Niklas Rönnberg},
  title     = {Musical sonification supports visual discrimination of color intensity},
  journal   = {Behaviour \& Information Technology (BIT)},
  volume    = {38},
  number    = {10},
  pages     = {1028--1037},
  year      = {2019},
  publisher = {Taylor \& Francis},
  doi       = {10.1080/0144929X.2019.1657952},
  url       = {https://doi.org/10.1080/0144929X.2019.1657952},
}

@incollection{blascheck2021characterizing,
  title={Characterizing glanceable visualizations: from perception to behavior change},
  author={Blascheck, Tanja and Bentley, Frank and Choe, Eun Kyoung and Horak, Tom and Isenberg, Petra},
  booktitle={Mobile Data Visualization},
  pages={151--176},
  year={2021},
  publisher={Chapman and Hall/CRC},
doi={https://doi.org/10.1201/9781003090823-5}
}

@article{moreira2024incidentalvisualization,
title = {Incidental visualizations: How complexity factors influence task performance},
journal = {Visual Informatics},
volume = {8},
number = {4},
pages = {85-96},
year = {2024},
issn = {2468-502X},
doi = {https://doi.org/10.1016/j.visinf.2024.10.005},
url = {https://www.sciencedirect.com/science/article/pii/S2468502X24000652},
author = {João Moreira and Daniel Mendes and Daniel Gonçalves}
}

@INPROCEEDINGS{blascheck2023parttowhole,
  title={Studies of Part-to-Whole Glanceable Visualizations on Smartwatch Faces}, 
  author={Blascheck, Tanja and Besançon, Lonni and Bezerianos, Anastasia and Lee, Bongshin and Islam, Alaul and He, Tingying and Isenberg, Petra},
  booktitle={IEEE Pacific Visualization Symp. (PacificVis)}, 
  year={2023},
  pages={187-196},
  doi={10.1109/PacificVis56936.2023.00028}}

@ARTICLE{liu2010mentalmodels,
  title={Mental Models, Visual Reasoning and Interaction in Information Visualization: A Top-down Perspective}, 
  author={Liu, Zhicheng and Stasko, John},
  journal={IEEE Trans. Visualization and Computer Graphics (TVCG)}, 
  year={2010},
  volume={16},
  number={6},
  pages={999-1008},
  doi={10.1109/TVCG.2010.177}
}

@ARTICLE{isenberg2013systematic,
  title={A Systematic Review on the Practice of Evaluating Visualization}, 
  author={Isenberg, Tobias and Isenberg, Petra and Chen, Jian and Sedlmair, Michael and Möller, Torsten},
  journal={IEEE Trans. Visualization and Computer Graphics (TVCG)}, 
  year={2013},
  volume={19},
  number={12},
  pages={2818-2827},
  doi={10.1109/TVCG.2013.126}}

@ARTICLE{richter2024scalability,
  author={Richer, Gaëlle and Pister, Alexis and Abdelaal, Moataz and Fekete, Jean-Daniel and Sedlmair, Michael and Weiskopf, Daniel},
  journal={IEEE Trans. Visualization and Computer Graphics (TVCG)}, 
  title={Scalability in Visualization}, 
  year={2024},
  volume={30},
  number={7},
  pages={3314-3330},
  doi={10.1109/TVCG.2022.3231230},
}

@article{park2025musicdynamics,
  title={Music dynamics visualization for music practice and education},
  author={Park, Eun Ji},
  journal={Multimedia Tools and Applications},
  pages={1--17},
  year={2025},
    month={jan},
  publisher={Springer},
    doi={10.1007/s11042-025-20637-0},
    url={https://doi.org/10.1007/s11042-025-20637-0}
}

@inproceedings{arai2023timtoshape,
title = {{TimToShape}: Supporting Practice of Musical Instruments by Visualizing Timbre with {2D} Shapes based on Crossmodal Correspondences},
author = {Arai, Kota and Hirao, Yutaro and Narumi, Takuji and Nakamura, Tomohiko and Takamichi, Shinnosuke and Yoshida, Shigeo},
year = {2023},
isbn = {9798400701061},
publisher = {ACM},
url = {https://doi.org/10.1145/3581641.3584053},
doi = {10.1145/3581641.3584053},
booktitle = {Proc. International Conf. Intelligent User Interfaces (IUI)},
pages = {850–865},
numpages = {16},
location = {Sydney, NSW, Australia},
}

@inproceedings{shu2025fretmate,
title = {{FretMate}: {ChatGPT}-Powered Adaptive Guitar Learning Assistant},
author = {Shu, Xin and Shi, Lei and Cheng, Jiacheng and Ouyang, Lingling and Chu, Mengdi and Shu, Xinhuan},
year = {2025},
isbn = {9798400713064},
publisher = {ACM},
url = {https://doi.org/10.1145/3708359.3712080},
doi = {10.1145/3708359.3712080},
booktitle = {Proc. International Conf. on Intelligent User Interfaces (IUI)},
pages = {715–726},
numpages = {12},
}

@inproceedings{neustaedter2012autobiographical,
author = {Neustaedter, Carman and Sengers, Phoebe},
title = {Autobiographical design in {HCI} research: designing and learning through use-it-yourself},
year = {2012},
isbn = {9781450312103},
publisher = {ACM},
url = {https://doi.org/10.1145/2317956.2318034},
doi = {10.1145/2317956.2318034},
booktitle = {Proc. Designing Interactive Systems Conf. (DIS)},
pages = {514–523},
numpages = {10},
}

@article{rosenblatt1956kde,
author = {Murray Rosenblatt},
title = {Remarks on Some Nonparametric Estimates of a Density Function},
volume = {27},
journal = {The Annals of Mathematical Statistics (AOMS)},
number = {3},
publisher = {Institute of Mathematical Statistics},
pages = {832-837},
year = {1956},
doi = {10.1214/aoms/1177728190},
URL = {https://doi.org/10.1214/aoms/1177728190}
}

@book{munzner2014bookvad,
  title={Visualization analysis and design},
  author={Munzner, Tamara},
  year={2014},
  publisher={CRC press},
doi={ttps://doi.org/10.1201/b17511},
isbn={9780429088902}
}

@ARTICLE{gleicher2023problemspace,
  author={Gleicher, Michael and Riveiro, Maria and von Landesberger, Tatiana and Deussen, Oliver and Chang, Remco and Gillman, Christina},
  journal={IEEE Computer Graphics and Applications (CG\&A)}, 
  title={A Problem Space for Designing Visualizations}, 
  year={2023},
  volume={43},
  number={4},
  pages={111-120},
  doi={10.1109/MCG.2023.3267213}}

@misc{miller2024melodyvis,
      title={{MelodyVis}: Visual Analytics for Melodic Patterns in Sheet Music}, 
      author={Matthias Miller and Daniel Fürst and Maximilian T. Fischer and Hanna Hauptmann and Daniel Keim and Mennatallah El-Assady},
      year={2024},
      eprint={2407.05427},
      archivePrefix={arXiv},
      primaryClass={cs.HC},
      doi={https://doi.org/10.48550/arXiv.2407.05427}, 
}

@book{aigner2011timevisbook,
    author = {Aigner, Wolfgang and Miksch, Silvia and Schumann, Heidrun and Tominski, Christian},
    title = {Visualization of Time-Oriented Data},
    publisher = {Springer},
    year = {2011},
    address = {London, UK},
    numpages = {286},
    isbn = {978-0-85729-078-6},
    doi = {10.1007/978-0-85729-079-3}
}

@ARTICLE{skreinig2023guitarhero,
  author={Skreinig, Lucchas Ribeiro and Kalkofen, Denis and Stanescu, Ana and Mohr, Peter and Heyen, Frank and Mori, Shohei and Sedlmair, Michael and Schmalstieg, Dieter and Plopski, Alexander},
  journal={IEEE Trans. Visualization and Computer Graphics (TVCG)}, 
 title={{guitARhero}: Interactive Augmented Reality Guitar Tutorials}, 
  year={2023},
  volume={29},
  number={11},
  pages={4676-4685},
  doi={10.1109/TVCG.2023.3320266},
  ISSN={1941-0506},
  month={Nov},}

@INPROCEEDINGS{maitz2023neuroadaptive,
  title={Towards Neuroadaptive Augmented Reality Piano Tutorials}, 
  author={Maitz, Florian and Ribeiro Skreinig, Lucchas and Kalkofen, Denis and Wriessnegger, Selina C.},
  booktitle={2023 IEEE International Conf. Metrology for eXtended Reality, Artificial Intelligence and Neural Engineering (MetroXRAINE)}, 
  year={2023},
  pages={450-455},
  doi={10.1109/MetroXRAINE58569.2023.10405829},
  month={Oct},}

@inproceedings{ng2015easytolearnpiano,
    title="Easy-to-Learn Piano: A Mobile Application for Learning Basic Music Theory and Piano Skill",
    author="Ng, S. C.
    and Lui, Andrew K. F.
    and Kwok, Alvin C. H.",
    booktitle="Technology in Education. Technology-Mediated Proactive Learning (ICTE)",
    year="2015",
    publisher="Springer Berlin Heidelberg",
    pages="103--112",
    isbn="978-3-662-48978-9",
    doi="10.1007/978-3-662-48978-9_10"
}

@incollection{stappers2014researchthroughdesign,
    title={Research through Design},
    booktitle = {The Encyclopedia of Human-Computer Interaction, 2nd Ed.},
    author={Stappers, Pieter J. and Giaccardi, Elisa},
    publisher={Interaction Design Foundation -- IxDF},
    year={2014},
    url={https://www.interaction-design.org/literature/book/the-encyclopedia-of-human-computer-interaction-2nd-ed/research-through-design}
}

@INPROCEEDINGS{heer2021fastkde,
  title={Fast \& Accurate {Gaussian} Kernel Density Estimation}, 
  author={Heer, Jeffrey},
  booktitle={IEEE Visualization Conf. (VIS)}, 
  year={2021},
  pusblisher={IEEE},
  pages={11-15},
  doi={10.1109/VIS49827.2021.9623323}
}

@string{Computer = "{IEEE} Computer" }

@string{Springer = "Springer-Verlag" }

@string{tvcg = {IEEE TVCG}}

@string{uist = {ACM UIST}}

@string{chi = {ACM CHI}}

@misc{heyen2022cellovis,
  title        = {Immersive Visual Analysis of Cello Bow Movements},
  author       = {Heyen, Frank and Kohler, Yannik and Triebener, Sebastian and Rigling, Sebastian and Sedlmair, Michael},
  year         = {2022},
  publisher    = {arXiv},
  doi          = {10.48550/ARXIV.2203.13316},
  url          = {https://arxiv.org/abs/2203.13316},
  copyright    = {Creative Commons Attribution Share Alike 4.0 International}
}

@inproceedings{heyen2023visual,
      title={Visual Overviews for Sheet Music Structure}, 
      author={Frank Heyen and Quynh Quang Ngo and Michael Sedlmair},
      year={2023},
      booktitle = {{Proc. International Society for 
                   Music Information Retrieval Conf. (ISMIR)}},
      publisher    = {ISMIR},
      doi = {https://doi.org/10.48550/arXiv.2308.06140}
}

@article{sedlmair2012design,
  title        = {Design Study Methodology: Reflections from the Trenches and the Stacks},
  author       = {Michael {Sedlmair} and Mariah {Meyer} and Tamara {Munzner}},
  year         = {2012},
  journal      = {IEEE Trans. Visualization and Computer Graphics (TVCG)},
  publisher    = {IEEE},
  volume       = {18},
  number       = {12},
  pages        = {2431--2440},
  doi          = {10.1109/TVCG.2012.213},
  issn         = {2160-9306}
}

@article{pousman2007casualvis,
  title        = {Casual Information Visualization: Depictions of Data in Everyday Life},
  author       = {Pousman, Zachary and Stasko, John and Mateas, Michael},
  year         = {2007},
  month        = {Nov},
  journal      = {IEEE Trans. Visualization and Computer Graphics (TVCG)},
  volume       = {13},
  number       = {6},
  pages        = {1145--1152},
  doi          = {10.1109/TVCG.2007.70541},
  issn         = {1941-0506}
}

@article{khulusi2020survey,
  title        = {A Survey on Visualizations for Musical Data},
  author       = {Khulusi, Richard and Kusnick, Jakob and Meinecke, Christofer and Gillmann, Christina and Focht, Josef and J{\"a}nicke, Stefan},
  year         = {2020},
  journal      = {Computer Graphics Forum (CGF)},
  doi          = {10.1111/cgf.13905},
  url          = {https://onlinelibrary.wiley.com/doi/abs/10.1111/cgf.13905},
  organization = {Wiley Online Library},
  eprint       = {https://onlinelibrary.wiley.com/doi/pdf/10.1111/cgf.13905}
}

@article{miller2022corpusvis,
  title        = {{CorpusVis}: Visual Analysis of Digital Sheet Music Collections},
    author = {Miller, Matthias and Rauscher, Julius and Keim, Daniel A. and El-Assady, Mennatallah},
    journal = {Computer Graphics Forum (CGF)},
    volume = {41},
    number = {3},
    pages = {283-294},
    doi = {https://doi.org/10.1111/cgf.14540},
    year = {2022}
}

@article{miller2022augmenting,
  title        = {Augmenting Digital Sheet Music through Visual Analytics},
  author       = {Miller, Matthias and Fürst, Daniel and Hauptmann, Hanna and Keim, Daniel A. and El-Assady, Mennatallah},
  year         = {2022},
  journal      = {Computer Graphics Forum (CGF)},
  volume       = {41},
  number       = {1},
  pages        = {301--316},
  doi          = {https://doi.org/10.1111/cgf.14436},
  url          = {https://onlinelibrary.wiley.com/doi/abs/10.1111/cgf.14436},
  eprint       = {https://onlinelibrary.wiley.com/doi/pdf/10.1111/cgf.14436}
}

@inproceedings{snydal2005improviz,
  title        = {{ImproViz}: Visual Explorations of {Jazz} Improvisations},
  author       = {Snydal, Jon and Hearst, Marti},
  year         = {2005},
  booktitle    = {Extended Abstracts on Human Factors in Computing Systems (CHI EA)},
  publisher    = {ACM},
  pages        = {1805–1808},
  doi          = {10.1145/1056808.1057027},
  isbn         = {1595930027},
  url          = {https://doi.org/10.1145/1056808.1057027},
  numpages     = {4}
}

@article{baur2010streams,
  title        = {The Streams of Our Lives: Visualizing Listening Histories in Context},
  author       = {Baur, Dominikus and Seiffert, Frederik and Sedlmair, Michael and Boring, Sebastian},
  year         = {2010},
  journal      = {IEEE Trans. Visualization and Computer Graphics (TVCG)},
  publisher    = {IEEE},
  volume       = {16},
  number       = {6},
  pages        = {1119--1128},
  doi          = {10.1109/TVCG.2010.206}
}

@inproceedings{watanabe2003brass,
	title        = {Brass: Visualizing scores for assisting music learning},
	author       = {Watanabe, Fumiko and Hiraga, Rumi and Fujishiro, Issei},
	year         = {2003},
	booktitle    = {Proc. International Computer Music Conf. (ICMC)},
    publisher    = {International Computer Music Association}
}

@article{huang2015personal:vis:and:va,
  title        = {Personal Visualization and Personal Visual Analytics},
  author       = {Huang, Dandan and Tory, Melanie and Adriel Aseniero, Bon and others},
  year         = {2015},
  journal      = {IEEE Trans. Visualization and Computer Graphics (TVCG)},
  volume       = {21},
  number       = {3},
  pages        = {420--433},
  doi          = {10.1109/TVCG.2014.2359887},
  issn         = {1941-0506},
}

@article{cantareira2016moshviz,
  title        = {{MoshViz}: A Detail+Overview Approach to Visualize Music Elements},
  author       = {Cantareira, Gabriel Dias and Nonato, Luis Gustavo and Paulovich, Fernando V.},
  year         = {2016},
  journal      = {IEEE Trans. Multimedia},
  volume       = {18},
  number       = {11},
  pages        = {2238--2246},
  doi          = {10.1109/TMM.2016.2614226},
  issn         = {1941-0077}
}

@inproceedings{bergstrom2007isochords,
  title        = {Isochords: Visualizing Structure in Music},
  author       = {Bergstrom, Tony and Karahalios, Karrie and Hart, John C.},
  year         = {2007},
  booktitle    = {Proc. Graphics Interface (GI)},
  publisher    = {ACM},
  pages        = {297–304},
  doi          = {10.1145/1268517.1268565},
  isbn         = {9781568813370},
  url          = {https://doi.org/10.1145/1268517.1268565},
  numpages     = {8}
}

@inproceedings{karolus2018emguitar,
  title        = {{EMGuitar}: Assisting Guitar Playing with Electromyography},
  author       = {Karolus, Jakob and Schuff, Hendrik and Kosch, Thomas and Wozniak, Pawe\l{} W. and Schmidt, Albrecht},
  year         = {2018},
  booktitle    = {Proc. Designing Interactive Systems Conf. (DIS)},
  publisher    = {ACM},
  pages        = {651–655},
  doi          = {10.1145/3196709.3196803},
  isbn         = {9781450351980},
  url          = {https://doi.org/10.1145/3196709.3196803},
  numpages     = {5}
}

@inproceedings{karolus2023eyepiano,
    author = {Karolus, Jakob and Sylupp, Johannes and Schmidt, Albrecht and Wo\'{z}niak, Pawe\l{} W.},
    title = {{EyePiano}: Leveraging Gaze For Reflective Piano Learning},
    year = {2023},
    publisher = {ACM},
    url = {https://doi.org/10.1145/3563657.3596065},
    doi = {10.1145/3563657.3596065},
    booktitle = {Proc. ACM Designing Interactive Systems Conf. (DIS)},
    pages = {1209--1223},
    numpages = {15},
    }

@inproceedings{marky2021letsfrets,
  title        = {Let’s Frets! Assisting Guitar Students During Practice via Capacitive Sensing},
  author       = {Marky, Karola and Wei\ss{}, Andreas and Matviienko, Andrii and Brandherm, Florian and Wolf, Sebastian and Schmitz, Martin and Krell, Florian and M\"{u}ller, Florian and M\"{u}hlh\"{a}user, Max and Kosch, Thomas},
  year         = {2021},
  booktitle    = {Proc. CHI Conf. Human Factors in Computing Systems (CHI)},
  publisher    = {ACM},
  isbn         = {9781450380966},
  url          = {https://doi.org/10.1145/3411764.3445595},
  doi          = {https://doi.org/10.1145/3411764.3445595},
  articleno    = {746},
  numpages     = {12}
}

@inproceedings{ariga2017strummer,
  title        = {Strummer: An Interactive Guitar Chord Practice System},
  author       = {Ariga, Shunya and Goto, Masataka and Yatani, Koji},
  year         = {2017},
  booktitle    = {International Conf. Multimedia and Expo (ICME)},
  publisher    = {IEEE},
  pages        = {1057--1062},
  doi          = {10.1109/ICME.2017.8019338},
  issn         = {1945-788X},
}

@inproceedings{loechtefeld2011guitar,
  title        = {{GuitAR}: Supporting Guitar Learning through Mobile Projection},
  author       = {L\"{o}chtefeld, Markus and Gehring, Sven and Jung, Ralf and Kr\"{u}ger, Antonio},
  year         = {2011},
  booktitle    = {CHI Extended Abstracts on Human Factors in Computing Systems (CHI EA)},
  publisher    = {ACM},
  pages        = {1447–1452},
  doi          = {10.1145/1979742.1979789},
  isbn         = {9781450302685},
  url          = {https://doi.org/10.1145/1979742.1979789},
  numpages     = {6}
}

@inproceedings{wang2021soloist,
  title        = {Soloist: Generating Mixed-Initiative Tutorials from Existing Guitar Instructional Videos Through Audio Processing},
  author       = {Wang, Bryan and Yang, Meng Yu and Grossman, Tovi},
  year         = {2021},
  booktitle    = {Proc. CHI Conf. Human Factors in Computing Systems (CHI)},
  publisher    = {ACM},
  isbn         = {9781450380966},
  url          = {https://doi.org/10.1145/3411764.3445162},
  doi          = {https://doi.org/10.1145/3411764.3445162},
  articleno    = {98},
  numpages     = {14}
}

@inproceedings{smith2008interactive,
  title        = {Interactive Software for Guitar Learning},
  author       = {Smith, Graeme and Johnston, Andrew},
  year         = {2008},
  booktitle    = {Australasian Computer Music Conf. (ACMC)},
  organization = {Australasian Computer Music Association}
}

@inproceedings{asahi2018toward:piano:support,
  title        = {Toward a High Performance Piano Practice Support System for Beginners},
  author       = {Asahi, Shota and Tamura, Satoshi and Sugiyama, Yuko and Hayamizu, Satoru},
  year         = {2018},
  booktitle    = {Asia-Pacific Signal and Information Processing Association Annual Summit and Conf. (APSIPA ASC)},
  pages        = {73--79},
  doi          = {10.23919/APSIPA.2018.8659463},
  issn         = {2640-0103},
}

@inproceedings{hori2019piano:hmm,
  title        = {Piano Practice Evaluation and Visualization by {HMM} for Arbitrary Jumps and Mistakes},
  author       = {Hori, Matsuto and Wilk, Christoph M. and Sagayama, Shigeki},
  year         = {2019},
  booktitle    = {Proc. Annual Conf. Information Sciences and Systems (CISS)},
  pages        = {1--5},
  doi          = {10.1109/CISS.2019.8692813},
}

@inproceedings{fober2007vemus,
  title        = {VEMUS - Feedback and Groupware Technologies for Music Instrument Learning},
  author       = {Fober, Dominique and Letz, St{\'e}phane and Orlarey, Yann},
  year         = {2007},
  booktitle    = {Sound and Music Computing Conf.},
  address      = {Lefkada, Greece},
  pages        = {117--123},
  url          = {https://hal.archives-ouvertes.fr/hal-02158814},
  hal_id       = {hal-02158814},
  hal_version  = {v1}
}

@article{deprisco2017understanding,
  title        = {Understanding the structure of musical compositions: Is visualization an effective approach?},
  author       = {Roberto De Prisco and Delfina Malandrino and Donato Pirozzi and Gianluca Zaccagnino and Rocco Zaccagnino},
  year         = {2017},
  journal      = {Information Visualization},
  volume       = {16},
  number       = {2},
  pages        = {139--152},
  doi          = {10.1177/1473871616655468},
  url          = {https://doi.org/10.1177/1473871616655468},
  eprint       = {https://doi.org/10.1177/1473871616655468}
}

@INPROCEEDINGS{ren2019forceguitar,
  title={Multi-Contact Force-Sensing Guitar for Training and Therapy}, 
  author={Ren, Zhiyi and Hsu, Chun-Cheng and Kocabalkanli, Can and Nguyen, Khanh and Iordachita, Iulian I. and Bastepe-Gray, Serap and Scott, Nathan},
  booktitle={IEEE SENSORS}, 
  year={2019},
  pages={1-4},
  doi={10.1109/SENSORS43011.2019.8956729},
  ISSN={2168-9229},
}

@inproceedings{fender2023pressurepick,
title = {{PressurePick}: Muscle Tension Estimation for Guitar Players Using Unobtrusive Pressure Sensing},
author = {Fender, Andreas and Witzig, Derek Alexander and M\"{o}bus, Max and Holz, Christian},
year = {2023},
isbn = {9798400701320},
publisher = {ACM},
url = {https://doi.org/10.1145/3586183.3606742},
doi = {10.1145/3586183.3606742},
booktitle = {Proc. Annual ACM Symp. on User Interface Software and Technology (UIST)},
articleno = {80},
numpages = {11},
}

@article{soszynski2016music,
  title={Music games as a tool supporting music education},
  author={Soszynski, Filip and Wo{\l}owski, Jakub and Stasiak, Bart{\l}omiej},
  journal={Computer Game Innovations},
  pages={116--132},
  year={2016}
}

@article{arsenault2008guitar,
  title={{Guitar Hero}: ``Not like playing guitar at all''?},
  author={Arsenault, Dominic},
  journal={Loading...},
  volume={2},
  number={2},
  year={2008}
}

@article{jenson2016explorationmusicvideogames,
   author = "Jenson, Jen and De Castell, Suzanne and Muehrer, Rachel and Droumeva, Milena",
   title = "So you think you can play: An exploratory study of music video games", 
   journal= "Journal of Music, Technology \& Education (JMTE)",
   year = "2016",
   volume = "9",
   number = "3",
   pages = "273-288",
   doi = "https://doi.org/10.1386/jmte.9.3.273_1",
   url = "https://intellectdiscover.com/content/journals/10.1386/jmte.9.3.273_1",
   publisher = "Intellect",
   issn = "1752-7074",
   type = "Journal Article",
  }

@article{omeara2016rocksmith,
  title={Rocksmith and the shaping of player experience},
  author={O’Meara, Daniel},
  journal={Music video games},
  pages={229--49},
  year={2016},
  publisher={Bloomsbury Academic}
}

@article{graham2018rockgod,
   author = "Graham, Kevin and Schofield, Damian",
   title = "Rock god or game guru: Using {Rocksmith} to learn to play a guitar", 
   journal= "Journal of Music, Technology \& Education",
   year = "2018",
   volume = "11",
   number = "1",
   pages = "65-82",
   doi = "https://doi.org/10.1386/jmte.11.1.65_1",
   url = "https://intellectdiscover.com/content/journals/10.1386/jmte.11.1.65_1",
   publisher = "Intellect",
   issn = "1752-7074",
   type = "Journal Article",
  }

@article{miller2009schizophonic, 
    title={Schizophonic Performance: {Guitar Hero, Rock Band}, and Virtual Virtuosity}, 
    volume={3}, 
    DOI={10.1017/S1752196309990666}, 
    number={4}, 
    journal={Journal of the Society for American Music (JSAM)}, 
    publisher={Cambridge University Press}, 
    author={Miller, Kiri}, 
    year={2009}, 
    pages={395–429}}

@ARTICLE{keebler2014fretlight,
	 TITLE={Shifting the paradigm of music instruction: Implications of embodiment stemming from an augmented reality guitar learning system},      
  AUTHOR={Keebler, Joseph R. and Wiltshire, Travis J. and Smith, Dustin C. and Fiore, Stephen M. and Bedwell, Jeffrey S.},   
	JOURNAL={Frontiers in Psychology},      
	VOLUME={5},           
	YEAR={2014},      
	  URL={https://www.frontiersin.org/articles/10.3389/fpsyg.2014.00471},       
	DOI={10.3389/fpsyg.2014.00471},      
	ISSN={1664-1078}
}

@incollection{shneiderman2003theeyeshaveitmantra,
  title        = {The Eyes Have It: A Task by Data Type Taxonomy for Information Visualizations},
  author       = {Ben Shneiderman},
  year         = {2003},
  booktitle    = {The Craft of Information Visualization},
  publisher    = {Morgan Kaufmann},
  series       = {Interactive Technologies},
  pages        = {364--371},
  doi          = {https://doi.org/10.1016/B978-155860915-0/50046-9},
  isbn         = {978-1-55860-915-0},
  url          = {https://www.sciencedirect.com/science/article/pii/B9781558609150500469}
}

@inproceedings{yamabe2011feedback,
  title        = {Feedback Design in Augmented Musical Instruments: A Case Study with an {AR} Drum Kit},
  author       = {Yamabe, Tetsuo and Asuma, Hiroshi and Kiyono, Sumire and Nakajima, Tatsuo},
  year         = {2011},
  booktitle    = {IEEE International Conf. Embedded and Real-Time Computing Systems and Applications (RTCSA)},
  publisher    = {IEEE},
  volume       = {2},
  pages        = {126--129},
  doi          = {10.1109/RTCSA.2011.27},
}

@article{deja2022surveyaugmentedpiano,
    title = {A Survey of Augmented Piano Prototypes: Has Augmentation Improved Learning Experiences?},
    author = {Deja, Jordan Aiko and Mayer, Sven and \v{C}opi\v{c} Pucihar, Klen and Kljun, Matja\v{z}},
    year = {2022},
    publisher = {ACM},
    volume = {6},
    number = {ISS},
    url = {https://doi.org/10.1145/3567719},
    doi = {10.1145/3567719},
    journal = {Proc. ACM Human-Computer Interaction (HCI)},
    articleno = {566},
    numpages = {28}
}

@article{moog1986midi,
  title        = {{MIDI}: Musical Instrument Digital Interface},
  author       = {Moog, Robert A.},
  year         = {1986},
  journal      = {Journal of the Audio Engineering Society},
  volume       = {34},
  number       = {5},
  pages        = {394--404},
}

@inproceedings{xia2018shift,
  title={{ShIFT}: A semi-haptic interface for flute tutoring},
  author={Xia, Gus and Jacobsen, Carter and Chen, Qianwen and Yang, Xingdong and Dannenberg, Roger},
  booktitle={International Conf. New Interfaces for Musical Expression (NIME)},
  year={2018},
  pages        = {162--167},
  publisher    = {Zenodo},
  month        = jun,
  doi          = {10.5281/zenodo.1302531},
  url          = {https://doi.org/10.5281/zenodo.1302531}
}

@inproceedings{pardue2019separating,
title = "Separating sound from source: sonic transformation of the violin through electrodynamic pickups and acoustic actuation",
author = "Pardue, {Laurel Smith} and Kurijn Buys and MIchael Edinger and Daniel Overholt and Andrew McPherson",
year = "2019",
month = mar,
day = "21",
pages = "278--283",
booktitle = "Proc. New Interfaces for Musical Expression Conf. (NIME)",
url = "https://www.nime.org/proceedings/2019/nime2019_paper053.pdf",
doi={https://doi.org/10.5281/zenodo.3672958}
}

\begin{IEEEbiography}[{\includegraphics[width=1in,height=1.25in,clip,keepaspectratio]{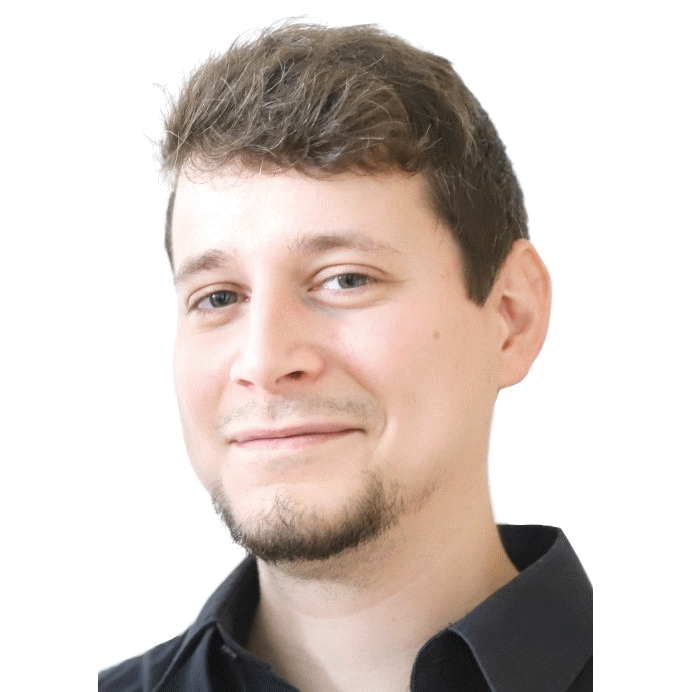}}]{Frank Heyen}
is currently working toward his PhD degree, for which he explores how visualization can support musicians in practicing their instruments.
His research includes visually encoding sheet music structure, using instruments as input for navigation, practice data visualization on screen or in augmented reality, and interactive composition through visualization of the output of generative models.
\end{IEEEbiography}

\begin{IEEEbiography}[{\includegraphics[width=1in,height=1.25in,clip,keepaspectratio]{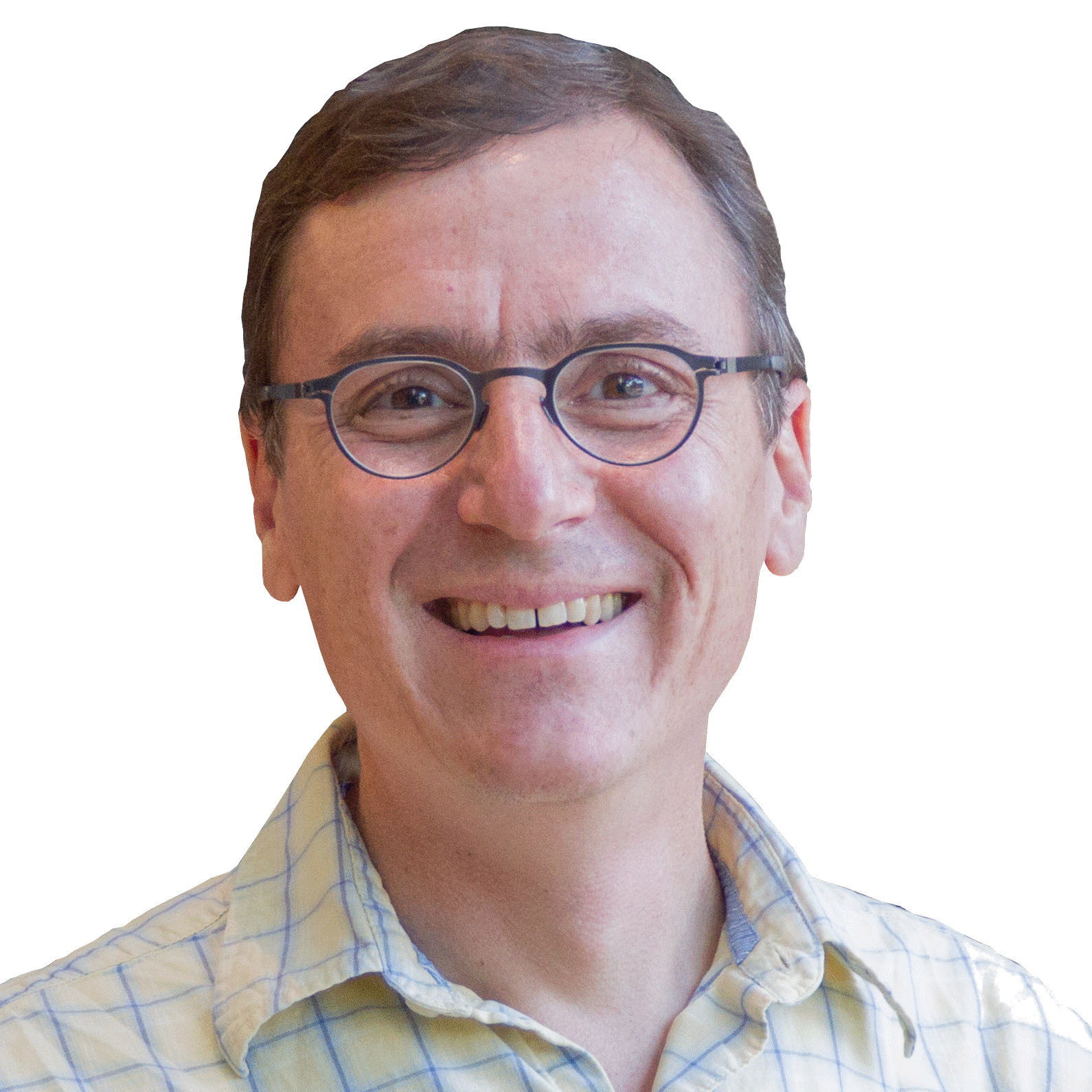}}]{Michael Gleicher}
(Senior Member, IEEE) is a Professor in the Department of Computer Sciences at the University of Wisconsin, Madison. 
He co-directs both the Visual Computing Laboratory and the Collaborative Robotics Laboratory at UW-Madison. 
He has been Papers Chair for EuroVis and Area Chair for IEEE VIS. 
Prior to joining the University, Prof. Gleicher was a researcher at Autodesk and Apple. 
He earned his Ph. D. in Computer Science from Carnegie Mellon University. 
He was a visiting researcher at INRIA Rhone-Alpes and holds a concurrent appointment as a Scholar at Amazon Robotics. This work is not associated with Amazon. 
Prof. Gleicher is an inductee to the IEEE Visualization Academy and an ACM Fellow. 
\end{IEEEbiography}

\begin{IEEEbiography}[{\includegraphics[width=1in,height=1.25in,clip,keepaspectratio]{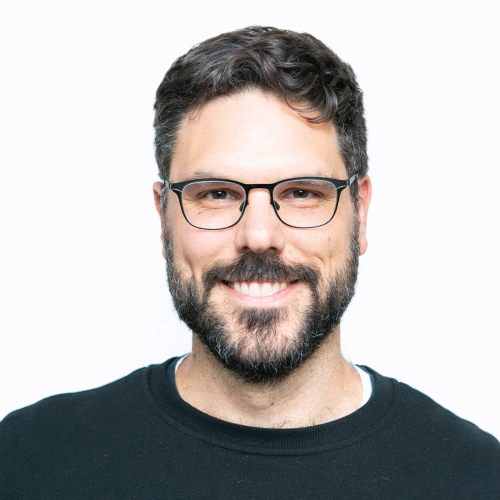}}]{Michael Sedlmair}
is a professor of Computer Science at the University of Stuttgart and leads the research group for Human-Computer Interaction. His research interests focus on visualization, augmented and virtual reality, and interaction design. Michael has been an active musician for more than 30 years, including seven years of teaching guitar at a music school.
\end{IEEEbiography}

\vfill

\end{document}